\documentclass[a4paper,onecolumn,11pt]
{quantumarticle}
\pdfoutput=1
\usepackage[dvipsnames]{xcolor}
\usepackage[utf8]{inputenc}

\usepackage{relsize}
\usepackage[english]{babel}
\usepackage[T1]{fontenc}
\usepackage{amsmath}
\usepackage{hyperref}

\usepackage{float}
\usepackage{braket}

\usepackage{tikz}
\usepackage{lipsum}

\usepackage[numbers,sort&compress]{natbib}

\begin{document}

\title{Optimal Control of thermally noisy quantum gates in a multilevel system}

\author{Aviv Aroch}
\email{aviv.aroch@mail.huji.ac.il}
\affiliation{ 
The Institute of Chemistry, The Hebrew University of Jerusalem, Jerusalem 9190401, Israel}
\orcid{0000-0002-0508-1992}

\author{Shimshon Kallush}
\email{shimshonk@hit.ac.il}
\affiliation{Sciences Department, Holon Academic Institute of Technology, 52 Golomb Street, Holon 58102, Israel}
\orcid{0000-0002-9036-5964}

\author{Ronnie Kosloff}
\email{kosloff1948@gmail.com}
\affiliation{ 
The Institute of Chemistry, The Hebrew University of Jerusalem, Jerusalem 9190401, Israel}
\orcid{0000-0001-6201-2523}

\begin{abstract}
\textcolor{black}{Quantum systems are inherently sensitive to environmental noise and imperfections in external control fields, which pose a significant challenge for the practical implementation of quantum technologies. These noise sources degrade the fidelity of quantum gates, making their mitigation a key requirement for realizing reliable quantum computing. In this study, we apply Optimal Control Theory (OCT) within a thermodynamically consistent Markovian framework to design high-fidelity quantum gates in the presence of thermal relaxation. Such a description is essential for realistic modeling and optimization of noisy quantum gates in near-term quantum technologies, where strong control fields and thermal environments act simultaneously.
Our approach combines OCT with a control-dependent dissipative generator derived from the non-adiabatic master-equation framework based on time dependent invariants of the free evolution. As a result the driving fields modify both the unitary and dissipative parts of the evolution. We implement the scheme for one- and two-qubit gates embedded in larger Hilbert spaces and compare direct-control and ancilla-assisted architectures.
Using logical-subspace-resolved diagnostics, we quantify how the optimized dynamics redistributes the dissipative action between logical and ancillary sectors in the model systems studied here. In particular, we show that ancilla-assisted control can reduce the effective thermal-noise burden on the logical subspace in the relevant parameter regime, while direct control remains the most effective route when available. High-precision propagation of the full open-system dynamics reveals substantial fidelity improvements, in some cases by orders of magnitude, while clarifying the limits of mitigation at large relaxation rates and temperatures.}
\end{abstract}

\noindent\textbf{Keywords:}
quantum optimal control;
open quantum systems;
Lindblad master equation;
thermal noise;
multilevel systems;
quantum gates.
\newpage

\section{Introduction}
\label{sec:intro}
The development of quantum technologies hinges critically on mitigating decoherence and the loss of quantum coherence arising from interactions between a system and its environment. As any physical quantum system is inevitably open, decoherence is a fundamental obstacle to scalable quantum information processing, quantum sensing, and other quantum applications \cite{preskill2018quantum,schlosshauer2019quantum}. While passive strategies such as isolation and material engineering seek to minimize environmental interactions, active noise-mitigation techniques offer dynamic alternatives that can adapt to specific operational conditions and system requirements. Developing control frameworks that remain valid under strong driving and realistic thermal conditions is, therefore, a central challenge for near-term quantum hardware.
Among these active strategies, Optimal Control Theory (OCT) has emerged as a robust mathematical framework for designing time-dependent external fields that steer quantum systems toward target states or unitary operations with high precision \cite{rice1992new,palao2003optimal,PhysRevLett.120.150401,koch2022quantum,muller2022one,turyansky2024inertial,george2025minimal,muller2022information,ansel2024introduction,khazali2026optimal}.

Initially developed for closed systems, OCT has since been extended to open quantum systems, where environmental interactions introduce dissipative and non-unitary dynamics \cite{bartana2001laser,koch2016controlling,fernandes2023effectiveness,malvetti2024reachability,ding2025universally,de2025fidelity,weidner2025robust,koch2016controlling,rodriguez2024optimal,cangemi2026control}. In this study, we apply OCT within a thermodynamically consistent formalism to mitigate thermal noise while executing quantum channels in particular quantum gate implementation. 

Any rational design of a quantum device requires the ability to simulate its operational protocol from first principles. Optimal-control studies share this requirement. Fault-tolerant quantum computing demands that the error per operation remain below a stringent threshold \cite{aharonov1997fault,marxer2025above,salatino2025noise}, imposing exceptionally high fidelity requirements on elementary quantum gates. Achieving this level of accuracy presents significant challenges for both the control strategy and the numerical methods used to simulate the device dynamics.

A key focus of this study is the suppression of Markovian thermal noise, characterized by rapid memoryless interactions between the system and its environment. This form of noise is particularly relevant for modern quantum devices, which operate at cryogenic temperatures and are susceptible to relaxation into thermal equilibrium. Such conditions exist in superconducting-based devices \cite{devoret2004superconducting}, but also for laser-cooled ion- or atomic-based systems \cite{kielpinski2002architecture,saffman2016quantum}. The first challenge of the study is to construct high-quality dynamical simulations of quantum channels in thermal environments.

To address time-dependent thermal noise, we use a thermodynamically consistent master equation that accurately captures the interplay between coherent dynamics and dissipative processes. Crucially, our formalism accounts for control-dependent dissipation, namely the fact that the external control field \( \varepsilon(t) \) alters not only the system’s unitary evolution but also the environment-induced dynamics by reshaping the instantaneous transition structure \cite{Dann2021quantumthermo,dann2022non}. 
For this task, we construct a new formulation that directly targets the equations of motion of quantum channels. This formalism is based on dynamical invariants or time-dependent constants of motion \cite{kaushal1993dynamical}.

Importantly, this invariant-based construction also clarifies the methodological distinction from the previous approximation strategy used in Ref.~\cite{kallush2022controlling}. In that work, the implementation of the inertial theory required repeated diagonalization of the driven generator \cite{dann2021inertial}. Here, the jump operators are instead propagated continuously in time using their smooth time dependence and an imaginary-time propagation scheme \cite{kosloff1994propagation}, thereby following the instantaneous time dependence of the map. This replaces the repeated diagonalization step by matrix--vector operations and provides a more direct and efficient route to the time-dependent dissipator.
\\
\\
With the framework established, the study focuses on OCT's ability to mitigate thermal noise.
We intend to address the following questions:
\begin{itemize}
    \item{How robust are OCT solutions for unitary dynamics to the addition of thermal noise?}
    \item{Does embedding the quantum gate in a larger Hilbert space reduce the influence of thermal noise?}
    \item{What are the conditions that OCT can mitigate thermal noise, and how does the temperature influence this task? }
    \item{Can we identify a mechanism of generalized cooling of a quantum channel? }
\end{itemize}

The foundations of the present study can be traced to the development of the non-adiabatic master equation \cite{dann2018time} and to advances in quantum control in thermal environments \cite{kallush2022controlling}. Building on these developments, the present work provides a comprehensive reformulation of the framework in terms of quantum channels. First, the dynamical equations of motion are constructed from dynamical invariants. Second, a novel propagator is introduced that directly addresses the challenge of time ordering. Finally, the full machinery of optimal control theory is employed to determine the control fields that realize the desired quantum operations.
This study is also complementary to our previous work \cite{aroch2024mitigating}, where we analyzed dephasing noise arising from imperfect control fields. 

The first example is designed to explore the potential benefits of embedding a qubit within a larger Hilbert space. Fault-tolerant quantum computation typically relies on logical qubits encoded across multiple elementary physical systems, with errors corrected at the gate level via redundancy \cite{roffe2019quantum}. In contrast, we investigate a more direct approach that operates at the dynamical level, exploiting the enlarged Hilbert space to enhance control and mitigate errors during the evolution itself.

First, we consider a single qubit coupled to an ancilla, in which the target gate is implemented indirectly via the ancilla in a three-level realization. Next, we extend this setting to include two- and three-ancilla states and investigate how the resulting enlargement of the Hilbert space affects the system's sensitivity to thermal relaxation. Finally, we examine the impact of supplementing the indirect scheme with a direct control field applied to the qubit, and assess its ability to further suppress noise and improve gate performance.

The second example is a
two-qubit entangling gate (controlled-\(iX\)). A central question throughout is how OCT improves gate fidelity and how the dissipative burden is distributed between the logical and ancillary sectors.

The novelty of the paper is the systematic combination of a thermodynamically consistent driven dissipator with gate-oriented OCT, together with a quantitative logical-subspace analysis of ancilla-assisted mitigation. This allows us to distinguish between direct-control improvement, ancilla-assisted reduction of effective logical-sector dissipation, and regimes in which thermal relaxation dominates regardless of optimization. For the task of analysis, we employ new tools
such as SU(3) visualization, channel purity, and channel energy loss, which indicate active cooling.

From a broader perspective, this work contributes to the ongoing development of thermodynamically consistent quantum control. As systems shrink in size and increase in complexity, thermodynamic principles such as entropy production, energy cost, and fluctuation-dissipation relations become increasingly relevant \cite{Fellous_Asiani_2023,morzhin2023control}. Integrating these concepts into quantum control frameworks yields not only more realistic models but also deeper insights into trade-offs among precision, efficiency, and robustness \cite{campbell2025roadmap}

The rest of this work is structured as follows: Section~\ref{sec:EOM} presents the open-system model, including the noise mechanisms and the thermodynamically consistent central equation. Section~\ref{sec:OCT} describes the OCT equations and the control-model architecture. Section~\ref{sec:results} presents numerical simulations of optimized gate operations and evaluates the performance of control strategies, including a logical-subspace-resolved analysis of ancilla-assisted mitigation. In Section~\ref{sec:discussion}, we interpret the results in the context of quantum thermodynamics and fault tolerance. Section~\ref{sec:conclus} summarizes the results and points out directions for future work.

\section{Equation of Motion}
\label{sec:EOM}
Our goal is to design an explicit time-dependent \emph{dynamical map} governing the reduced-system dynamics, aiming to create a target map that accounts for environmental noise.
The objective can be described by a completely positive trace-preserving map (CPTP)\cite{kraus1971general}, or more generally, an open-system quantum channel,
\begin{equation}
  \hat  \rho(0) \mapsto \hat \rho(T) = \Phi_T[\hat \rho(0)] .
\end{equation}
This channel $\Phi_T$ maps states represented 
as density operators $\hat \rho$ in a time duration $T$.
The task of quantum control is to find time-dependent control fields such that the generated map $\Lambda_T$ approximates a target operation $\Phi_T$ with maximal fidelity while respecting the dynamical and experimental constraints of the device.

A prerequisite for  constructing a control simulation
is a dynamical equation of motion (EOM) for the channel:
\begin{equation}
    \frac{d}{dt} \Lambda_t = {\cal L} \Lambda_t
\end{equation}
which initial condition $\Lambda_0={\bf  I}$. ${\cal L}$
is the generator formally: ${\cal L} =\dot \Lambda \Lambda^{-1}$ assuming the map has an inverse.

Here, $\Phi_T$ may represent a target unitary gate, a noisy quantum
channel \cite{holevo2012quantum}, or a prescribed superoperator in Liouville space.
This map-based formulation generalizes the more common state-to-state control
approach \cite{koch2016controlling} and is particularly suited to quantum
information processing tasks, where operations must be, implemented reliably on
arbitrary input states \cite{palao2003optimal}.

For isolated quantum systems the simulation task is 
considerably simplified. The map becomes unitary:
$\Lambda_t = {\cal U}_t$, the EOM becomes:
\begin{equation}
    \frac{d}{dt} {\cal U} = -i [\hat H_t,{\cal U} ]
    \label{eq:vonNeumann_isolated}
\end{equation}
where the dynamics is generated by the Hamiltonian $\hat H_t = \hat H_0 + \hat H_c(t)$ which
is partitioned into a static drift Hamiltonian $\hat H_0$
and a time-dependent control part $H_c(t)$.
For a finite quantum system, a control solution is guaranteed. The setup is completely controllable, meaning any unitary map can be generated,
provided the commutator algebra generated from $\hat H_0$
and $H_c(t)$ forms a complete operator base in Hilbert space \cite{huang1983controllability}.

For open quantum devices, obtaining the generator ${\cal L}_t$ is inherently more difficult. It can be partitioned into unitary and dissipative parts:
${\cal L}= -i[ \hat H, \bullet]+ {\cal D}$.
Thermodynamic consistency requires that the dissipative and the unitary parts are linked.
The time dependence of the control Hamiltonian
imposes dynamical symmetry such that the dissipative part is also time dependent.

\subsection{Global Unitary Evolution}

\begin{figure}
    \centering
    \includegraphics[width=0.85\textwidth]{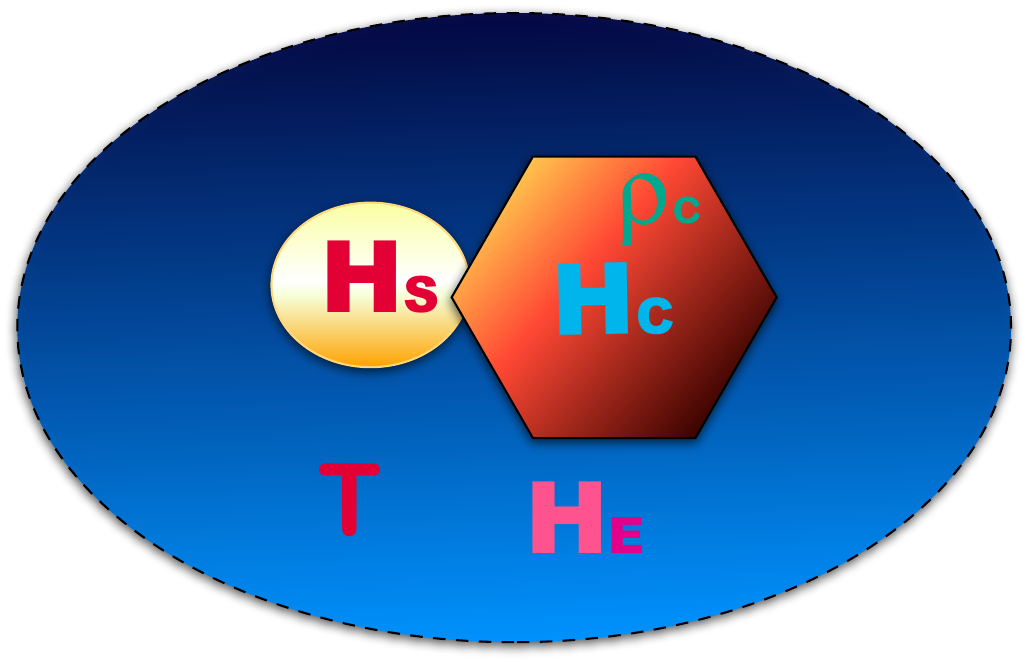}
    \caption{Global unitary dynamics. The primary system is coupled to a controller embedded in a thermal environment. The controller initial state $\hat \rho_C(0)$ is the source of coherence.}
    \label{fig:sysbath}
\end{figure}

Open quantum system dynamics is based on a 
global, unitary description of the joint evolution of the system (S), controller
(C), and environment (E).
The environment is assumed to be large and stationary in thermal equilibrium.
The controller is large and nonstationary, responsible
for coherence that drives the map, Cf. Fig. \ref{fig:sysbath}.

The system is coupled to the controller via \(\hat{H}_{SC}\), and the composite
device (S+C) is coupled to the environment \cite{Dann2021quantumthermo}:
\begin{equation}
    \hat{H}_G
    = \hat{H}_S^0 + \hat{H}_C + \hat{H}_{SC}
    + \hat{H}_{DE} + \hat{H}_E
    \equiv \hat{H}_D + \hat{H}_{DE} + \hat{H}_E ,
\end{equation}
where \(\hat{H}_{DE}\) describes the device--environment interaction
\cite{dann2021quantum}.

The full system evolves unitarily according to
the Liouville--von Neumann equation,
\begin{equation}
    \frac{d}{dt} \hat \rho_G(t)
    = -\frac{i}{\hbar}
    \left[ \hat{H}_G,\hat \rho_G(t) \right],
\end{equation}
with the formal solution
\begin{equation}
\begin{aligned}
\hat{\rho}_G(t)
= e^{-i[\hat H_G,\bullet]t}\hat{\rho}_G(0)
\end{aligned}
\end{equation}

We assume the systems are initially uncorrelated: the primary system, the controller, and the environment,
\begin{equation}
   \hat \rho_G(0)
    = \hat \rho_S(0) \otimes \hat \rho_C(0) \otimes \hat \rho_E ,
    \label{eq:initial_state_global}
\end{equation}
where the environment is assumed to be in a stationary thermal equilibrium state at temperature
\(T\),
\begin{equation}
   \hat \rho_E
    = \frac{e^{-\hat{H}_E / k_B T}}{Z_E},
    \qquad
    Z_E
    = Tr_E \left\{ e^{-\hat{H}_E / k_B T} \right\}.
    \label{eq:thermal_environment}
\end{equation}
The controller is initialized in a non-stationary state satisfying
\(
[\hat \rho_C(0), \hat{H}_C] \neq 0
\),
which is essential for coherent control.
Under global unitary evolution, coherence is a constant of motion
\cite{dann2023unification}, and control is achieved by redistributing coherence
from the controller to the system degrees of freedom.

The reduced state of the system is obtained by tracing out the controller and
environmental degrees of freedom,
\begin{equation}
   \hat \rho_S(t)
    = Tr_E \!\left\{
        Tr_C \!\left[
            \hat{U}_G(t,0)\,
            \big(
                \hat \rho_S(0)\otimes \hat \rho_C(0)\otimes\hat \rho_E
            \big)\,
            \hat{U}_G^\dagger(t,0)
        \right]
    \right\}.
    \label{eq:reduced_dynamics_trace}
\end{equation}
This transformation defines a CPTP dynamical map \(\Lambda_t\) acting on the
system,
\begin{equation}
  \hat  \rho_S(t) = \Lambda_t[\hat \rho_S(0)] ,
    \label{eq:dynamical_map_hilbert}
\end{equation}
which fully characterizes the reduced, generally non-unitary dynamics of the
system and is guaranteed to be CPTP due to its reduction from a unitary
evolution on the enlarged Hilbert space \cite{kraus1971general}.

Provided the map $\Lambda_t$ has an inverse, then
it has a time-local differential generator ${\cal L}$. 
To obtain the explicit form of ${\cal L}$, two additional assumptions are employed. The first assumes an isothermal partition between the environment, the system, and the controller. Such a partition allows energy transfer between the two 
components without accumulating energy on the interface: $    [\hat H_{DE},\hat H_G] =0 $ \cite{dann2021quantum}.
This condition implies that the unitary free propagator of the device ${\cal U}_D$ and its dynamical map $\Lambda_D$ commute $[\Lambda_D,{\cal U}_D]=0$. This symmetry condition is a 
consequence of the environment being stationary.
The second condition is on the controller. It
is assumed to be large relative to the system
so that the back-action of the system on the controller dynamics can be ignored \cite{dann2023unification}.

The free propagator can be expressed in the interaction representation with
respect to the controller Hamiltonian as
\begin{equation}
\mathcal{U}_D(t)
=
{\mathcal{U}}_C(t)\,\tilde{\mathcal{U}}_{SC}(t),
\qquad
\mathcal{U}_C(t)=e^{-\frac{i}{\hbar}[\hat H_C,\bullet]t}
\label{eq:interac}
\end{equation}
where \(\tilde{\mathcal{U}}_{SC}(t)\) is the propagator in the interaction picture
generated by \([\tilde{H}_{SC} + \hat{H}_S,\bullet]\), and \(\tilde{H}_{SC}(t)=\hat U_C^\dagger(t)\,\hat H_{SC}\,\hat U_C(t)\) is the system--controller coupling
in the interaction representation with respect to \(\hat H_C\),

The system propagator is obtained by tracing over the controller degrees of
freedom. Assuming that the controller is large compared to the system, so that
its dynamics is predominantly generated by \(\hat{H}_C\) and back-action can be
neglected, the free system propagator becomes
\begin{equation}
    \mathcal{U}_S(t)
    =
    Tr_C\!\left\{
       \hat \rho_C(t)\,\tilde{\mathcal{U}}_{SC}(t)
    \right\},
\end{equation}
where $\rho_C(t)={\cal U}_C(t) \hat \rho_C(0)$.
This procedure yields an effectively driven system Hamiltonian of the form \cite{dann2021quantum}:
\begin{equation}
    H(t) = H_0 + \sum_{\alpha} \epsilon_\alpha(t) H_\alpha .
\end{equation}
where $H_0$ is the drift Hamiltonian and $H_c=\sum_{\alpha} \epsilon_\alpha(t) H_\alpha$ is the control Hamiltonian. And $\epsilon_\alpha (t)$ are the semiclassical control fields.

The map  obeys a time-local master equation
\begin{equation}
    \frac{d}{dt}\Lambda(t) = \mathcal{L}_t[\Lambda(t)] ,
    \label{eq:general_master}
\end{equation}
where the generator is decomposed into coherent and dissipative parts,
\begin{equation}
    \mathcal{L}_t[\bullet]
    = -\frac{i}{\hbar}[H(t),\bullet] + \mathcal{D}_t[\bullet] .
    \label{eq:Liouvillian_general}
\end{equation}
For Markovian dynamics, $\mathcal{D}_t$ has the Gorini--Kossakowski--Lindblad--Sudarshan form \cite{dann2018time,dann2021quantum}:
\begin{equation}
    \mathcal{D}_t[\bullet]
    = \sum_k \gamma_k(t)
    \left(
    L_k(t)\bullet L_k^\dagger(t)
    -\frac{1}{2}\left\{L_k^\dagger(t)L_k(t),\bullet\right\}
    \right) .
    \label{eq:GKLS}
\end{equation}
The dependence of the rates $\gamma_k(t)$ and jump operators $L_k(t)$ on the drive is essential whenever the control modifies the system's instantaneous transition structure.

The symmetry condition imposed by the free propagator of the device and its map can be reduced under the semiclassical approximation to
\begin{equation}
    \left[\widetilde{\mathcal{U}}_S(t),\Lambda_t\right]=0,
    \label{eq:covariance}
\end{equation}
where $\mathcal{U}_S(t)$ denotes the free system propagator generated by the controlled Hamiltonian, 
where $\Lambda_t$ is the reduced dissipative map. Equation~\eqref{eq:covariance} means that the dissipator must respect the dynamical symmetry generated by the driven system. 

We can now move one step further and employ the symmetry condition between the free propagator and the dissipative generator:
\begin{equation}
    [{\cal L}(t),\widetilde{\mathcal{U}}_S(t)]=
    [{\cal D}(t),\widetilde{\mathcal{U}}_S(t)]=0
\end{equation}
With this condition, a common operator basis can be defined for both the free evolution and the dissipative generator ${\cal D}$. As a result, the jump operators, which are eigenoperators of ${\cal D}$, can be calculated as eigenoperators of the free evolution $\widetilde{\mathcal{U}}_S(t)$.

This construction leads naturally to a drive-dependent dissipator. The transition frequencies and jump operators are extracted from the instantaneous or invariant structure of the driven dynamics.

\subsection{Invariant-Based Free Unitary Evolution and Thermodynamic Consistency}
\label{sec:invariant_unitary}

Our task is to obtain the common time-dependent operator basis of the free propagator and the dissipator. The chosen approach is first to obtain the eigenoperator basis of the free time-dependent propagator ${\cal U}(t)$.

The eigenoperators of ${\cal U}(t)$ are divided into two classes: invariants of
the free motion and jump operators.
In general, an invariant of dynamics is defined as a time-dependent constant of motion. For an observable
\(\hat{A}\)
\cite{kaushal1993dynamical,levy2018noise,cangemi2023control,PRXQuantum.5.040346,jin2025universal,takahashi2025krylov,boubakour2025dynamical,cangemi2026control},
we obtain:
\begin{equation}
 \frac{\partial}{\partial t}\hat{A}
    + i[\hat{H}_S(t),\hat{A}]
    = 0 ,
    \label{eq:invariant}
\end{equation}
Eq. (\ref{eq:invariant}) is solved using a complete set of operators \(\{\hat{B}\}\) forming a closed Lie algebra \cite{martinez2026symdyn}:
\begin{equation}
    [\hat{B}_i, \hat{B}_j] = \sum_k \mathcal{C}_{ij}^k \hat{B}_k,
    \label{eq:liealgebra}
\end{equation}
where \(\mathcal{C}_{ij}^k\) are the structure constants of the algebra.
The time-dependent Hamiltonian is expanded in this basis as
\begin{equation}
    \hat{H}_S(t) = \sum_l h_l(t) \hat{B}_l.
    \label{eq:hamilh}
\end{equation}
Expanding the invariant in the same basis,
\[
    \hat{A}(t) = \sum_n c_n(t)\,\hat{B}_n,
\]
and inserting into Eq.~\eqref{eq:invariant} leads to the propagation equation
\begin{equation}
    \frac{\partial}{\partial t} \hat{A} =
    \sum_n \dot{c}_n \hat{B}_n
    = -i \sum_{l,n,k} \mathcal{C}_{ln}^k\,h_l(t)\,c_n(t)\,\hat{B}_k .
    \label{eq:inprop}
\end{equation}
In matrix form, the evolution equation for the coefficient vector
\(\vec{c}(t)\) becomes \cite{cangemi2023control}
\begin{equation}
    \frac{d}{dt} \vec{c}(t) = \boldsymbol{\mathcal{M}}(t)\,\vec{c}(t),
    \label{eq:m-matrix}
\end{equation}
with matrix elements
\(
\boldsymbol{\mathcal{M}}_{kn}(t) = -i \sum_l \mathcal{C}_{ln}^k\,h_l(t)
\).
In order to obtain eigenoperators of the free evolution
we explicitly choose the initial conditions to become the spectral decomposition of the stationary Hamiltonian \(\hat{H}_S(0)\),
\begin{equation}
    \hat{H}(0) = \sum_j \epsilon_j\,|j \rangle \langle j|,
    \label{eq:hamilproj}
\end{equation}
Therefore the initial invariants operators are \(\hat{A}_j(0) = |j \rangle \langle j|\), generating \(N\) initial conditions.

The time-dependent jump operators, which are transient eigenoperators
forming transition operators between pairs of invariants \(\hat{A}_i(t)\) and \(\hat{A}_j(t)\). They are obtained by solving the eigenvalue problem in Liouville space
\begin{equation}
    [\hat{A}_i(t) - \hat{A}_j(t), \hat{F}_{ij}(t)] = -2\,\hat{F}_{ij}(t),
    \label{eq:eigeninver}
\end{equation}
where the eigenoperators \(\hat{F}_{ij}(t)\) with the lowest eigenvalue \(-2\) become the time-dependent jump operators. The corresponding instantaneous jump frequency \(\omega_{ij}(t)\) for each \(\hat{F}_{ij}(t)\) is obtained by inserting the operator into the time-dependent Heisenberg equation:
\begin{equation}
    i[\hat{H}_S(t), \hat{F}_{ij}(t)] - \frac{\partial}{\partial t} \hat{F}_{ij}(t)
    = \omega_{ij}(t)\,\hat{F}_{ij}(t).
    \label{eq:frequency}
\end{equation}
The eigenfrequencies \(\omega_{ij}(t)\) determine the kinetic coefficients \(\gamma_k(t)\) and enforce time-dependent detailed balance. Initially $\omega_{ij}(0)=\epsilon_i -\epsilon_j$.

To illustrate the structure of these frequencies, we plot in Fig.~\ref{fig:bohr_frequency} the instantaneous transition (Bohr) frequencies \(\omega_{ij}(t) \) for all two-level sub-manifolds of the driven three-level system.

The time-dependent sets of invariants \(\{\hat{A}_i(t)\}\) and jump operators
\(\{\hat{F}_{ij}(t)\}\) together form a complete orthogonal operator basis in
Liouville space. In this basis, the free evolution superoperator
\(\boldsymbol{\mathcal{U}}(t)\) is diagonal:
\begin{align*}
\boldsymbol{\mathcal{U}}(t)\,\hat{A}_{i}(t) &= \hat{A}_{i}(t), \\
\boldsymbol{\mathcal{U}}(t)\,\hat{F}_{ij}(t) &= e^{i \phi_{ij}(t)}\,\hat{F}_{ij}(t),
\end{align*}
where \(\phi_{ij}(t) = \int_0^t \omega_{ij}(t')\,dt'\) and
\(\omega_{ij}(t)\) are the instantaneous transition frequencies induced by the
external driving. Therefore, the free propagator becomes:
\begin{equation}
\boldsymbol{\mathcal{U}}^{N^2}(t) =
\begin{bmatrix}
\underbrace{
\begin{matrix}
1 &  & \cdots &  & 0 \\
 & 1 & \cdots &  &  \\
\vdots & \vdots & \ddots & \vdots &  \\
 &  & \cdots & 1 &  \\
0 &  &  &  & 1 
\end{matrix}
}_{\text{size } N} & 
0 \\[10pt]
0 & 
\underbrace{
\begin{matrix}
e^{i \phi_{11}(t)} &  &  &  \\
 & e^{i \phi_{ij}(t)} &  &  \\
 &  & \ddots &  \\
 &  &  & e^{i \phi_{lk}(t)}
\end{matrix}
}_{\text{size } N^2 - N}
\end{bmatrix}.
\end{equation}
We can  now check for the eigenoperators of
${\cal L}$ and ${\cal D}$ Eq. (\ref{eq:Liouvillian_general})(\ref{eq:GKLS}).
For the jump operators:
\begin{eqnarray}
    {\cal L}(t)\hat{F}_{ij}(t)= \left(~~i \omega_{ij}(t)-\frac{1}{2}(\gamma_i(t) +\gamma_j(t))\right)\hat{F}_{ij}(t)\\
        {\cal L}(t)\hat{F}_{ji}(t)= \left(-i \omega_{ij}(t)-\frac{1}{2}(\gamma_i(t) +\gamma_j(t))\right)\hat{F}_{ji}(t)
\end{eqnarray}
Since the spectrum of the invariants $\{\hat A_j \} $ is degenerate the eigenoperators
of ${\cal D}$ become a linear combination of the set.

For thermodynamic consistency with an environment with temperature $T$
the kinetic coefficients should satisfy the detailed-balance relations,
\begin{align}
    \gamma_{ij}=\kappa\Gamma^{\uparrow}_{i\leftarrow j}(t)
    &\propto J\!\left(\omega_{ij}(t)\right)n_T\!\left(\omega_{ij}(t)\right), \\
    \gamma_{ji}=\kappa\Gamma^{\downarrow}_{j\leftarrow i}(t)
    &\propto J\!\left(\omega_{ij}(t)\right)
    \left[n_T\!\left(\omega_{ij}(t)\right)+1\right],
    \label{eq:thermal_rates}
\end{align}
where $\kappa$ is a scaling factor,
$J(\omega)$ is the bath spectral density and $n_T(\omega)=\frac{1}{e^{\hbar \omega/{k T}} -1}$.
Conservation laws constrain the structure of the dynamics in the open-system formulation.

\subsubsection{Quantum System in a Thermal Bath}
\label{section: relaxation model}

\begin{figure} 
\includegraphics[width=0.47\textwidth,angle=-90]{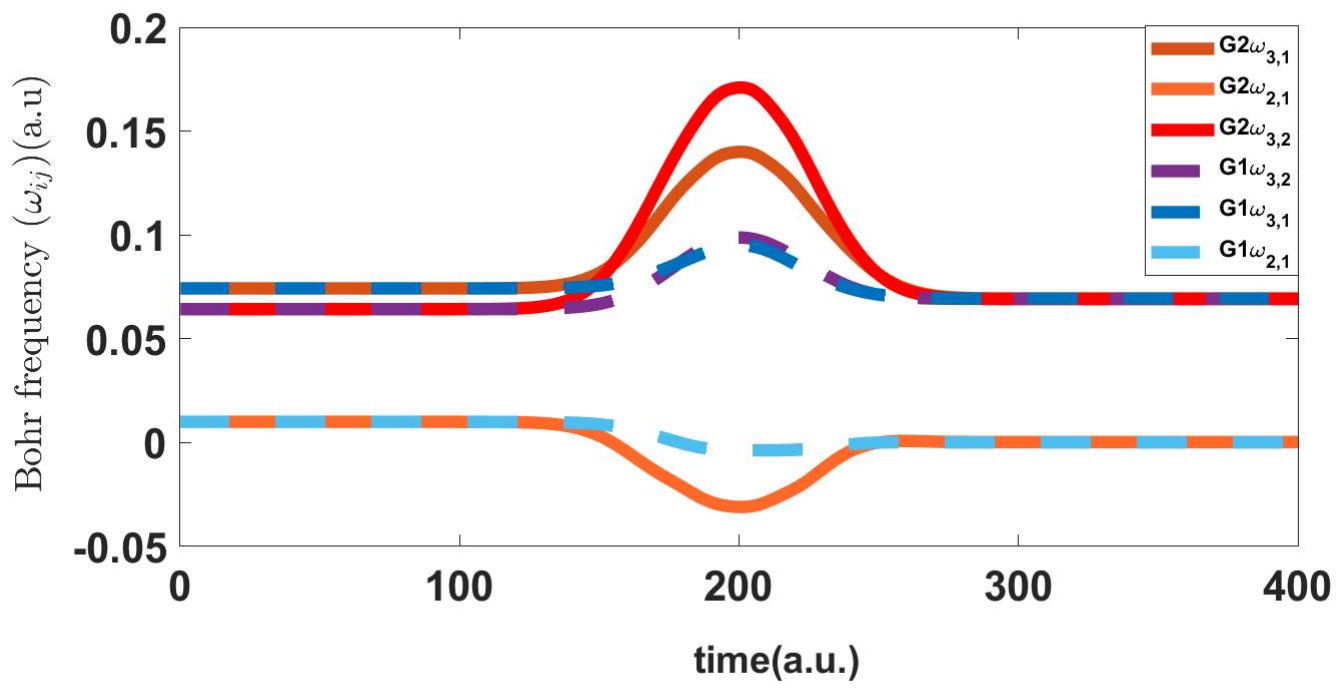}
    
    \caption{Instantaneous transition (Bohr) frequencies \(\omega_{ij}(t)\) (a.u.) for all two-level sub-manifolds of the driven three-level system, plotted versus time (a.u.). We show here two different cases with the same drift Hamiltonian but different control protocols (dashed and solid). these are the frequencies used in fig. (\ref{fig:2_case_Control_1Qbit}) for the single qubit Hadamard gate.}
    \label{fig:bohr_frequency}
\end{figure}

To illustrate the structure of these frequencies, we plot in Fig.~\ref{fig:bohr_frequency} the instantaneous transition (Bohr) frequencies \(\omega_{ij}(t) \) for all two-level sub-manifolds of the driven three-level system.
The frequencies are obtained from two optimized control solutions 
“G1” and “G2” described in subsection \ref{subsec:single-qubit}. The protocols have the same $\hat H_0$ but different $\varepsilon_{1,2}(t)$), yielding distinct instantaneous spectra $\{\omega_{ij}(t)\}$.

Notably, around the pulse center (\(t \approx 200\) a.u.), the drive significantly reshapes the spectrum: specific gaps are enlarged while others are compressed, leading to near-degenerate levels. These near-degeneracies create "hot spots" prone to non-adiabatic leakage and thermally assisted transitions.

A striking indicator of non-adiabaticity is observed in the fact that several Bohr frequencies do not revert to their initial values by the pulse's conclusion, indicated by \(\omega_{ij}(t_{\mathrm{final}}) \neq \omega_{ij}(0)\). In a purely adiabatic process, the spectrum would return to its original configuration. Therefore, \(\omega_{ij}(t)\) serves as a direct spectral indicator, revealing when the system is most susceptible to both coherent and thermal noise, and illustrating how different gate designs (G1 vs. G2) affect the structure of these gaps.

Finally, collecting the ingredients above, we obtain the Gorini, Kossakowski, Lindblad, and Sudarshan (GKLS) master equation
\cite{lindblad1976generators,gorini1976completely} in the invariant basis. The
Lindbladian \({\cal L}=-i[H(t),\bullet]+\mathcal{D}(t)\) in the time-local GKLS form becomes:
\begin{equation}
\mathcal{D}(t)=\sum_{i \ne j}
\Gamma_{ij}(t)\!\left[
\hat F_{ij}(t)\bullet \hat F_{ij}^\dagger(t)
-\tfrac{1}{2}\{\hat F_{ij}^\dagger(t)\hat F_{ij}(t),\bullet\}
\right],
\label{eq:NAME-GKLS}
\end{equation}
where \(\Gamma_{ij}= \Gamma_{i\leftarrow j}^{\uparrow}(t)\) and
\(\Gamma_{ji}= \Gamma_{j\leftarrow i}^{\downarrow}(t)\).

The use of invariants as the foundation for constructing jump operators is motivated by their ability to capture the long-time dynamical structure while preserving coherence and symmetry. Invariants provide a bridge between the unitary evolution dictated by \(\hat{H}_S(t)\) and the dissipative contributions introduced by the environment, ensuring that the resulting master equation is consistent with the underlying physical constraints. By defining the jump operators through the propagation of invariants, the dissipative terms induce transitions only within the system’s natural eigen-structure, thereby respecting fundamental conservation laws. This approach offers a systematic way to incorporate complex environmental interactions without resorting to ad hoc approximations, leading to a more accurate and computationally efficient description of open-system dynamics.

Practically, this construction also replaces the previous approximation strategy of Ref.~\cite{kallush2022controlling}, eliminating the computationally intensive diagonalization step previously required in the inertial theory \cite{dann2021inertial}. We utilize the time continuity of the jump operators to propagate them efficiently from one time step to the next using imaginary-time propagation \cite{kosloff1994propagation}, so that each update reduces to simple matrix–vector multiplications. Moreover, the invariance-based framework naturally lends itself to extensions beyond the Markovian limit \cite{dann2022non}, providing the flexibility needed to treat memory effects and strong coupling in more general non-Markovian settings.

\subsection{Controller-noise models}
\label{sec:controller_noise}

In addition to environmental noise, the pulse generator or controller 
is an independent source of noise.
The worst case is fast Markovian noise. Amplitude noise is defined as \cite{gorini1976completely}:
\begin{equation}
    \mathcal{D}_{A}[\bullet]
    = -\gamma_A \epsilon^2(t)[H_c,[H_c,\bullet]] .
    \label{eq:amp_noise}
\end{equation}
A corresponding phase or timing noise leads to \cite{feldmann2006quantum}:
\begin{equation}
    \mathcal{D}_{P}[\bullet]
    = -\gamma_P [H(t),[H(t),\bullet]] .
    \label{eq:phase_noise}
\end{equation}
Control mitigating exclusively controller noise was studied previously \cite{aroch2023employing}.
Can OCT address both sources of noise simultaneously?

\section{Numerical Propagation}
\label{sec:propagation}

To solve the Liouville–von Neumann equation, achieving high-fidelity control of quantum gates requires highly accurate and efficient numerical propagators. 
The challenge is to treat generators that are explicitly time-dependent and possess complex eigenvalue spectra. For this purpose, we adapted the semi-global propagator \cite{schaefer2017semi} to operate within the Liouville vector space.

For a driven open system, the propagator is partitioned into a time-independent and time-dependent generator:
\begin{equation}
\begin{aligned}
\frac{d}{d t} \boldsymbol{\Lambda}(t)
={\boldsymbol{\mathcal{L}}}(t) \boldsymbol{\Lambda}(t) =\left(
{\boldsymbol{\mathcal{L}}}_0+{\boldsymbol{\mathcal{L}}}_t\right)\boldsymbol{\Lambda}(t)
\end{aligned}
\label{eq:tlio-1}
\end{equation}
The time-independent part typically is generated by the drift Hamiltonian
$\boldsymbol{\mathcal{L}}_0=-i[\hat H_0,\bullet]$.  The time-dependent
component is composed from the control Hamiltonian and the time-dependent dissipator:
${\cal L}_t=-i[\hat H_t,\bullet]+{\cal D}_t$.

For a time-independent Lindbladian ${\boldsymbol{\mathcal{L}}}_0$ the formal solution of the dynamics $\frac{d}{dt}\boldsymbol{\Lambda}(t)= {\boldsymbol{\mathcal{L}}}_0 \boldsymbol{\Lambda}(t) $,  the propagator  becomes:
\begin{equation}
    \boldsymbol{\Lambda}(t)=e^{{\boldsymbol{\mathcal{L}}}_0 t}
    \label{time-independent Lindbladian}
\end{equation}
with the initial conditions $\boldsymbol{\Lambda}(0)={\bf I}$. 
Employing the partition into a time-dependent and time-independent part, a formal solution of Eq. (\ref{eq:tlio}) can be formulated as an integral equation:
\begin{equation}
    \boldsymbol{\Lambda}(t)=e^{{\boldsymbol{\mathcal{L}}}_0 t}+
    \int_0^t e^{{\boldsymbol{\mathcal{L}}}_0 (t-\tau)}{\cal L}_{\tau}\boldsymbol{\Lambda}(\tau) d\tau
    \label{integral-solution-2}
\end{equation}
Eq. (\ref{integral-solution-2}) will form the basis for the numerical approximation. 
The  idea is to solve the integral equation iteratively with a time step $\Delta t$. Then, one can concatenate the propagators and obtain the total evolution from $t=0$ to $t=\tau$ by
\begin{equation}
\boldsymbol{\Lambda}(t) \approx \prod_{j=1}^{N_t} {\mathcal Q }_j(\Delta t) 
\label{Prudoct-rule}
\end{equation}
where ${\mathcal Q }_j(\Delta t)$ is the propagator for $t$ to $t+\Delta t$ and $t=j \Delta t$. To speed up the convergence, we interpolate the solution from one time step to the next.

The equation of motion of ${\mathcal Q}_j(\Delta t)$ 
also requires a method to solve the integral equation. 
\begin{equation}
    {\mathcal Q}_j(t+\Delta t)=e^{{\boldsymbol{\mathcal{L}}}_0 \Delta t}\left( 1+
    \int_t^{t_j+\Delta t} e^{{\boldsymbol{\mathcal{L}}}_0 (-\tau)} {\cal L}_{t_j+\tau}{\mathcal Q}_j(t_j+\tau)d\tau \right)
    \label{integral-solution-1}
\end{equation}
In the first term $e^{{\boldsymbol{\mathcal{L}}}_0 \Delta t}$ the generator ${\cal L}_0$ is time independent.
This suggests a polynomial expansion of the exponent. Since the eigenvalues of ${\cal L}_0$
can be complex, we employ a Newtonian interpolation polynomial in the complex plane \cite{ashkenazi1995newtonian}. We now face the problem of how to incorporate the full integral equation into a polynomial form in ${\cal L}_0$. The first step is to extrapolate the propagator
on the rhs from the previous iteration
${\cal L}_{t_j+\tau}{\mathcal Q}^{n-1}_j(t_j+\tau)$.
The problem now becomes how to evaluate the 
integral by a polynomial expansion and incorporate it with the free propagation 
$e^{{\boldsymbol{\mathcal{L}}}_0 \Delta t}$.
To facilitate a solution of the integral
we expand ${\cal L}_{t_j+\tau}{\mathcal Q}_j(t_j+\Delta t)$ 
in a power series in $\tau$
\begin{equation}
    {\cal L}_{t_j+\tau}{\mathcal Q}^{n-1}_j(t_j+\tau)=\sum_k^M {\cal V}_k \frac{\tau^k}{k!}
\end{equation}
This expansion is obtained by sampling the operator function ${\cal L}_{q}{\mathcal Q}_j(q)$ at Chebyshev sampling points of the second kind, which include the endpoints. This choice of sampling points enables high accuracy and facilitates the extrapolation from one time step to the next. The values of the function on the sampling points are obtained  by a linear transformation of the expansion terms
${\cal L}_{q}{\mathcal Q}_j(q)$.

The transformation of the integral into a polynomial
is based on the integral $\int x^k e^{\alpha x}  dx =  \frac{1}{\alpha}x^ke^{\alpha x}-\frac{k}{\alpha}\int x^{k-1}e^{\alpha x} dx$ which leads to a recurrence relation where the high powers of $\tau$ are added successfully to the lower ones \cite{tal2012new}.

\begin{equation}
\Lambda_t = \sum_{j=0}^{m-1} \frac{t^j}{j!} {\cal R}_j + f_m({\cal L}_0,\tau){\cal R}_m,
\end{equation}
where ${\cal R}_j$ satisfy the recurrence relation
\begin{align}
{\cal R}_0 &= \bf I, \\
{\cal R}_j &= {\cal L} {\cal R}_{j-1} + {\cal V}_{j-1}, \qquad 1 \leq j \leq m
\end{align}
and the function of the generator $z={\cal L}_0$:
\begin{equation}
f_m(z,t)=
\begin{cases}
\dfrac{1}{z^m}
\left(
e^{zt}
-
\displaystyle\sum_{j=0}^{m-1}
\frac{(zt)^j}{j!}
\right),
& z \neq 0, \\[12pt]
\dfrac{t^m}{m!},
& z = 0.
\end{cases}
\end{equation}

The final step is to expand the function of the free propagator
$f({\cal L}_0,\tau)$ by a polynomial. 
Since the Liouvillian is non-Hermitian, the eigenvalue domain becomes complex; then the Newton or Arnoldi approach should be adopted \cite{kosloff1994propagation,arnoldi1951principle,lehoucq1996deflation}. 
The Krylov subspace is generated from an initial vector ${\cal G}$ and a generator ${\cal L}$:
\begin{equation}
    \mathcal{K}_m({\cal L},{\cal G})=\operatorname{span}\{{\cal G},{\cal L}{\cal G},{\cal L}^2 {\cal G},\ldots,{\cal L}^{m-1} {\cal G}\} .
\end{equation}
The Arnoldi procedure produces an orthonormal basis $\bf V_m$ and projected matrix $\bf H_m$ such that
\begin{equation}
  {\cal L}\bf V_m \approx V_m H_m .
\end{equation}
The operator function is then approximated by
\begin{equation}
    f({{\cal L}_0},\tau){\cal G}
    \approx
    {\bf V_m} f({ {\bf H_m},\tau}) {\bf V_m^\dagger} {\cal G} .
    \label{eq:krylov_exp}
\end{equation}
Additional details can be found in Appendix \ref{app:prop}.

\section{Optimal Control Theory (OCT) for quantum channels}
\label{sec:OCT}

Our primary objective is to execute a quantum gate while mitigating the impact of noise.
To achieve this task, we utilize Optimal Control Theory (OCT) to compute effective control fields. The external fields, denoted $\left\{\varepsilon(t) \right\}$, guide the system's dynamics from an initial state to a desired final state. An upper-level objective is to generate a quantum map $\Phi(T)=\Lambda(T)$ that can execute the desired gate \cite{palao2003optimal,khodjasteh2010arbitrarily,carolan2023robustness,aroch2024mitigating,george2025minimal}. 

For the description, we employ a complete basis set of orthogonal operators and use them to vectorize
Liouville space (Appendix \ref{sec:veccing}).
The equation of motion governing this map is expressed as:
\begin{equation}
\frac{d\boldsymbol{\Lambda}(t)}{dt}= {\boldsymbol{\mathcal{L}}}(t)\,\boldsymbol{\Lambda}(t),
\qquad
\boldsymbol{\Lambda}(0)=\boldsymbol{\mathcal I}. 
   \label{Vecor-Louivile}
\end{equation}
where $\boldsymbol{\mathcal{L}}(t)$ is the generator of the dynamics or in Liouville space ${ \boldsymbol{\mathcal  L}}$
represented as a matrix.

The time dependence of the generator $\boldsymbol{\mathcal{L}}(t)$ is determined by the control Hamiltonian $\hat H_c$ and control fields $\left\{\vec u(t)\right\}$:
\begin{equation}
\label{eq:Hcont}
    \hat H_c =\sum_l u_l(t) \hat H_l
\end{equation}
The objective is to determine the optimal driving fields $\left\{\vec u(t)\right\}$ that induce a desired transformation ${\mathcal{O}}$ in $t=\tau$. This task requires mapping a complete set of operators $\{ \hat{\boldsymbol{A}} \}$ as generated by the target map ${\mathcal{O}}$. We define the fidelity $F$ as the metric of reaching the objective \cite{palao2003optimal,aroch2024mitigating,george2025minimal,goerz2014optimal,schulte2011optimal,fernandes2023effectiveness}:
\begin{equation}
F
= \frac{1}{N^2}\mathrm{Tr}\!\left\{\mathcal{O}^{\dagger}\,\Lambda(\tau)\right\}
\label{eq:fidelity}
\end{equation}
where $N$ is the size of the Hilbert space.
The fidelity is normalized such that $0 \le F \le 1$.

In the framework of OCT, the control task is cast as the maximization of an objective functional \cite{petruhanov2023quantum}  $\mathcal{J}_{\max}  \propto F$ where:
\begin{equation}
\mathcal{J}_{\max}
= \mathrm{Tr}\!\left\{\mathcal{O}^{\dagger}\,\Lambda(\tau)\right\}
= \sum_j \operatorname{tr}\!\left[\big(\mathcal{O}\hat{ A}_j\big)^{\!\dagger}\,\big(\Lambda(\tau)\hat{ A}_j\big)\right]
\label{eq:Jmax}
\end{equation}
Here, \(\Lambda(\tau)\) is the superoperator (map) at time \(\tau\), acting on operators as \(\hat X \mapsto \Lambda(\tau)\hat X\). The set \(\{\hat{ A}_j\}\) is an orthonormal operator basis (Hilbert–Schmidt inner product).  We reserve \(\mathrm{Tr}\{\cdot\}\) for the (matrix) trace over superoperator representations, and \(\operatorname{tr}\{\cdot\}\) for the usual Hilbert-space trace. To simplify the derivation, we only include a single control field $u_1(t)=\varepsilon(t)$.

Two constraints are added. First, the dynamics must satisfy the Liouville equation
\begin{equation}
\dot{\Lambda}(t)=\mathcal{L}(t)\,\Lambda(t),\qquad \Lambda(0)=\mathcal{I},
\label{eq:Liouville-eq}
\end{equation}
enforced by a Lagrange-multiplier superoperator \(\Upsilon(t)\):
\begin{equation}
\mathcal{J}_{\mathrm{con}}
= \int_0^\tau \mathrm{Tr}\!\left\{\big(\dot{\Lambda}(t)-\mathcal{L}(t)\Lambda(t)\big)\,\Upsilon(t)\right\}\,dt.
\label{eq:Jcon}
\end{equation}
Second, we penalize control energy,
\begin{equation}
\mathcal{J}_{\mathrm{pen}}=\lambda \int_0^\tau\frac{|\varepsilon(t)|^2}{s(t)}\,d\tau,
\label{eq:Jpen}
\end{equation}
with \(\lambda>0\) and a smooth shape \(s(t)\) (here Gaussian).

The total functional is
\begin{equation}
\mathcal{J}_{\mathrm{Tot}}=\mathcal{J}_{\max}+\mathcal{J}_{\mathrm{con}}+\mathcal{J}_{\mathrm{pen}},
\label{eq:Jtot}
\end{equation}
and stationarity \(\delta \mathcal{J}_{\mathrm{Tot}}=0\) with respect to \(\Upsilon,\Lambda,\epsilon\) yields:

\begin{enumerate}
\item \textbf{Forward map propagation:} the Liouville equation \eqref{eq:Liouville-eq} with \(\Lambda(0)=\mathcal{I}\).

\item \textbf{Adjoint (backward) propagation:}
\begin{equation}
\dot{\Upsilon}(t)=\mathcal{L}^{\dagger}(t)\,\Upsilon(t),
\qquad
\Upsilon(\tau)=\mathcal{O}^{\dagger},
\label{eq:adjoint-eq}
\end{equation}
where \(\mathcal{L}^{\dagger}\) is the adjoint with respect to the Hilbert–Schmidt inner product. In matrix form,
\(\dot{\tilde{\boldsymbol Y}}(t)={\boldsymbol{\mathcal L}}^{\dagger}(t)\,\tilde{\boldsymbol Y}(t)\), with \(\tilde{\boldsymbol Y}(\tau)=\tilde{\boldsymbol{\mathcal O}}^{\dagger}\).

\item \textbf{Field update (gradient/Krotov step):} for a Hamiltonian part linear in the field, \(\mathcal{L}_{H_c}(t)
=\varepsilon(t)\,\mathcal{L}'_{H_c}\), and the dissipator depends quadratically on the field. Defining $\gamma \mathcal{L}'_D=\frac{1}{2}\frac{\partial}{\partial \varepsilon^2}\mathcal{L}_D$
leads to the update term:
\begin{equation}
\begin{aligned}
\Delta \varepsilon(t)
&= -\,\frac{s(t)}{2}\,
\operatorname{Im}\!
\left[
\frac{\mathrm{Tr}\!\left\{\Upsilon(t)\,\mathcal{L}'_c\,\Lambda(t)\right\}}
{\lambda+\gamma_A\,\mathrm{Tr}\!\left\{\Upsilon(t)\,\mathcal{L}'_D\,\Lambda(t)\right\}}
\right] 
\end{aligned}
\label{eq:update}
\end{equation}
where \(\tilde{\boldsymbol{\mathcal H}}'_c \equiv \partial \tilde{\boldsymbol{\mathcal L}}/\partial \epsilon\) at fixed state. 
In most cases studied, \(\lambda \gg \gamma_A\,\big|\mathrm{Tr}\{\cdot\}\big|\), therefore the denominator correction can be neglected, yielding the standard Krotov-type step\cite{rodriguez2024optimal}:
\begin{equation}
\begin{aligned}
\Delta \epsilon^{(k)}(t)
&= -\,\frac{s(t)}{2\lambda}\,
\operatorname{Im}\!\left[\,
\mathrm{Tr}\!\left\{\Upsilon^{(k-1)}(t)\,\mathcal{L}'_c\,\Lambda^{(k)}(t)\right\}
\right] 
\end{aligned}
\label{eq:update-krotov}
\end{equation}
\item \textbf{Reconstructing the equation of motion.}
With the updated field $\varepsilon$ recalculating the invariants Eq. (\ref{eq:invariant}) and ${\cal L}_D(t)$ Eq. (\ref{eq:NAME-GKLS}).
\end{enumerate}

Iterations proceed until the desired objective value (equivalently, absolute infidelity \(1-\mathcal{J}_{\max}\)) is reached.

\subsection{GRAPE based OCT}
\label{sec:grape}

An alternative OCT scheme is GRAPE. 
The method can  result in high fidelity solutions \cite{khaneja2005optimal,goodwin2016modified,petruhanov2022optimal,petruhanov2023grape}. The starting point is the control Hamiltonian Eq. (\ref{eq:Hcont}.
Nevertheless, the GRAPE control fields differ significantly from the ones obtained by the Krotov scheme \cite{palao2003optimal}.
GRAPE treats the control amplitudes as optimization variables and uses gradients of the objective with respect to the discretized pulse. Let $u_j$ be the control amplitude on interval $j$. The channel is written as a product of short-time propagators,
\begin{equation}
\Phi(T)=\boldsymbol{\Lambda}(T) \approx \prod_{j=1}^{N_t} {\mathcal Q }_j(\Delta t) 
\nonumber
\end{equation}

The derivative is
\begin{equation}
    \frac{\partial {\Lambda}(T)}{\partial u_j}
    = {\cal Q}_{N}\cdots{\cal Q}_{j+1}
    \frac{\partial{\cal Q}_{j}}{\partial u_j}
    {\cal Q}_{j-1}\cdots{\cal Q}_{1} .
    \label{eq:grape_derivative}
\end{equation}
The key numerical issue is the evaluation of $\partial{\cal Q}_{j}/\partial u_j$. If the actual propagation uses an Arnoldi--Chebyshev semi-global method, then the derivative should be computed consistently with that propagation rather than assuming a simple first-order exponential derivative. A matched sensitivity equation can be propagated together with the state:
\begin{align}
    \frac{d}{dt}\left(\frac{\partial{\Phi}}{\partial u_j}\right)
    &= {\cal L}(t)\frac{\partial{\Phi}}{\partial u_j}
    + \frac{\partial{ \cal L}(t)}{\partial u_j}{\Lambda}(t) .
    \label{eq:sensitivity}
\end{align}
This equation provides a gradient compatible with the chosen equation of motion and propagation scheme.
The addition of OCT employing GRAPE is aimed at increasing the family of high-fidelity solutions
to ensure that our observations are not inferred from a single rare example.

\subsection{{Control of quantum gates}}
\label{subsec:control}

Control solutions for quantum gates 
are obtained by applying OCT ( section \ref{sec:OCT}) \cite{palao2003optimal,grace2007optimal,schulte2011optimal,goerz2014optimal,sauvage2022optimal,zindorf2025efficient,gautier2025optimal,sun2025control,de2025fidelity}.
The generator of the dynamical map  is chosen as:
\begin{equation}
{\boldsymbol{\mathcal L}}=
{\boldsymbol{\mathcal L}}_{H_0}
+
{\boldsymbol{\mathcal L}}_{H_c}
+
{\boldsymbol{\mathcal L}}_{H_{uc}}
+
{\boldsymbol{\mathcal L}}_D .
    \label{total dynamical map}
\end{equation}
where ${\boldsymbol{\mathcal H}}_0$ generates the drift, ${\boldsymbol{\mathcal H}}_c$ generates the time dependent control ${\boldsymbol{\mathcal H}}_{uc}$ is a small static control employed to break symmetry and $\boldsymbol{\mathcal{L}}_D$ generates the thermal noise. 

Optimal control is employed first
to obtain the desired unitary gate
without dissipation.
This solution serves as a reference
for studying the effect of noise on fidelity. At this point, optimal control 
is used again, including the dissipation, to search for control fields that mitigate
the impact of noise.

\begin{figure}

    \centering
    \includegraphics[width=0.5\linewidth]{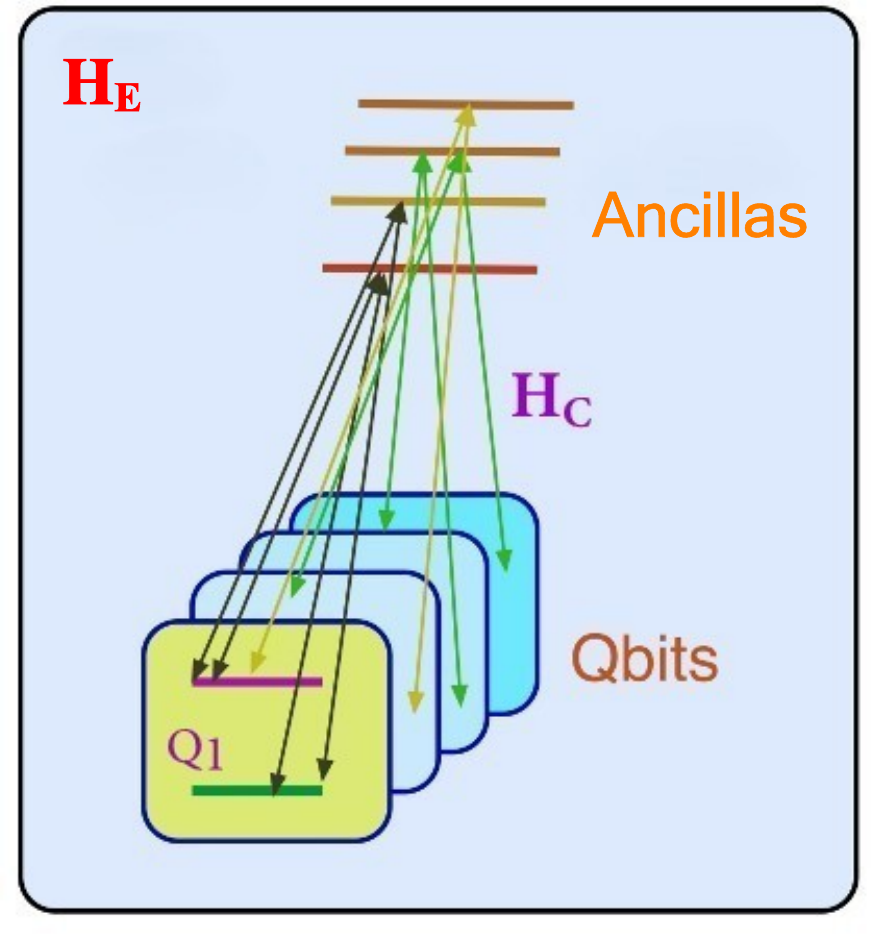}
    \caption{Schematic illustration of the system architecture. The primary quantum system consists of isolated qubits \(Q_1, Q_2, \ldots, Q_N\). The control field acts indirectly via a set of ancilla modes \(a_1, a_2, \ldots, a_N\), resembling Raman transitions. All components, including both qubits and ancillas, are coupled to a shared thermal environment \(H_E\), which introduces decoherence and dissipation. This structure underlies the control framework analyzed in this study, in which noise is mitigated, and coherence is preserved through optimal control strategies tailored to this topology.
}
    \label{fig: System-ancila-bath}
\end{figure}

We consider a driven quantum system that evolves unitarily and is controlled only through the ancilla degree of freedom, as shown in Fig. \ref{fig: System-ancila-bath}.
The inclusion of ancillas increases the dimension of the Hilbert space, which directly impacts the ability to obtain high-fidelity solutions in the noiseless limit. This added complexity manifests as a more rugged control landscape in the enlarged Hilbert space \cite{bondar2022globally}. As a result, the search for high-fidelity solutions depends crucially on the initial random guess.

To facilitate the search and break
possible hidden symmetry, we
introducing a small, static,  direct interaction ${\boldsymbol{\mathcal H}}_{uc}$. With this addition, we successfully realize the target unitary and uncover a small family of pulse solutions that differ in their temporal structures yet yield comparable figures of merit.

When thermal noise is factored in, indirect control becomes demanding; the convergence rate of fidelity is very slow, with minimal improvement over the noisy system. Notably, even a slight direct drive on the logical transition significantly improves
the error mitigation through OCT.

An additional investigation 
considers a two-qubit system with direct control. We choose to implement the controlled-\(iX\) two-qubit entangling gate also studied in ref. \cite{aroch2024mitigating}. The dimension of the gate
is four, embedded in a Hilbert space of 16.
Our analysis focuses on the fidelity of this gate as we vary the environment's temperature and rates.
The effect can then be compared
to controller noise studied previously \cite{aroch2024mitigating}.



We aimed to investigate the influence of thermal noise on our gate performance. In all of our simulations, we began with a closed system. By employing Optimal Control Theory (OCT), we identified the pulse that produces a unitary map over a full basis, achieving an infidelity of approximately \(5 \times 10^{-5}\). This pulse, along with its associated infidelity, served as our initial reference for all simulations of the open quantum system exposed to thermal noise. The infidelity \(IF_U\) was used as a normalization metric for the losses induced by noise in these simulations.

\subsection{Model Hamiltonian and System Architecture}
\label{sec:model}

\subsubsection{Operator Basis: Gell-Mann Matrices}

To model quantum systems of dimension \( d \geq 3 \), we use the Gell-Mann matrix basis \( \{G_k\}_{k=1}^{d^2 - 1} \) \cite{hioe1985gell,martinez2026symdyn}, which spans the Lie algebra \({su}(d) \). These matrices are traceless, Hermitian, and satisfy:
\begin{equation}
\mathrm{Tr}[G_j G_k] = 2 \delta_{jk}, \qquad G_k^\dagger = G_k.
\end{equation}

In the three-level case (\(d = 3\)), we use:
\begin{align}
\nonumber
G_1 &=
\begin{pmatrix}
0 & 1 & 0 \\
1 & 0 & 0 \\
0 & 0 & 0
\end{pmatrix}. &
G_3 &=
\begin{pmatrix}
1 & 0 & 0 \\
0 & -1 & 0 \\
0 & 0 & 0
\end{pmatrix}, \\
\nonumber
G_8 &= \frac{1}{\sqrt{3}}
\begin{pmatrix}
1 & 0 & 0 \\
0 & 1 & 0 \\
0 & 0 & -2
\end{pmatrix}, &
G_4 &=
\begin{pmatrix}
0 & 0 & 1 \\
0 & 0 & 0 \\
1 & 0 & 0
\end{pmatrix}, \\
G_6 &=
\begin{pmatrix}
0 & 0 & 0 \\
0 & 0 & 1 \\
0 & 1 & 0
\end{pmatrix}.
\end{align}
Here, \(G_1\) is the direct single qubit transitions  \(G_3\) and \(G_8\) define the energy structure, while \(G_4\) and \(G_6\) mediate transitions between the qubit and ancilla.

\subsubsection{Single Qubit with One Ancilla (\texorpdfstring{$d = 3$}{d = 3})}

The three-level system consists of a qubit \((|0\rangle, |1\rangle)\) and one ancilla \(|a\rangle\). The generating Hamiltonian is:
\begin{equation}
\hat{H}(t) = \hat{H}_0 + \hat{H}_c(t) + \hat{H}_{uc}(t),
\end{equation}
with the drift Hamiltonian:
\begin{equation}
\hat{H}_0 = \frac{\omega}{2} G_3 + \frac{4\omega}{2\sqrt{3}} G_8,
\end{equation}
ensuring a gap \(\omega\) between qubit states and a detuning \(4\omega\) for the ancilla.

The controlled and uncontrolled interactions are:
\begin{align}
\hat{H}_c(t) &= \epsilon_4(t) G_4 + \epsilon_6(t) G_6, \\
\hat{H}_{uc}(t) &= \epsilon_4^{uc}(t) G_1.
\end{align}

\subsubsection{One Qubit with Two or Three Ancillas}

For \(N = 2\) or \(3\), the Hilbert-space dimension is \(d = N + 2\). The Hamiltonian structure generalizes as:
\begin{equation}
\hat{H}_0 = \frac{\omega}{2} {G}_{\text{qubit}} + \sum_{j=1}^N \Delta_j\,{G}^{(a_j)}, \quad \Delta_j = 4j \omega,
\end{equation}
where

\[
\begin{aligned}
{G}_{\text{qubit}}
&= \mathrm{diag}(1, -1, 0, \dots, 0), \\
{G}^{(a_j)}
&= \mathrm{diag}(0, 0, \dots, \underbrace{1}_{a_j}, \dots, 0).
\end{aligned}
\]

The interaction Hamiltonians are:
\begin{align}
\hat{H}_c(t) &= \sum_{j=1}^{N} \left[ a_j\varepsilon(t)\, G_4^{(j)} + b_j\epsilon_{6,j}(t)\, G_6^{(j)} \right], \\
\hat{H}_{uc}(t) &= \sum_{j=1}^{N} \left[ c_j\epsilon_{uc}(t)\, G_4^{(j)} + d_j\epsilon_{uc}(t)\, G_6^{(j)} \right],
\end{align}
with
\[
G_4^{(j)} = |0\rangle\langle a_j| + |a_j\rangle\langle 0|, \quad
G_6^{(j)} = |1\rangle\langle a_j| + |a_j\rangle\langle 1|.
\]
The numerical procedures are based on
vectorizing Liouville space described in Appendix \ref{sec:veccing} and the high accuracy propagator described in Appendix \ref{app:prop}.

\section{OCT Results}
\label{sec:results}

Our noise-mitigation study proceeds in three stages.
First, we define the target dynamical map.
For a single qubit augmented by ancilla levels, we choose the Hadamard gate
\(\boldsymbol{\mathcal{O}}_H\) acting on the logical subspace.
In the Liouville-space representation, the corresponding superoperator takes
the block-structured form
\begin{equation}
\boldsymbol{\mathcal{O}}_H=
\left(
    \begin{array}{ccccc|cccc}
1 &  1 & 0 &  1 &  1 & 0 & 0 & 0 & 0\\
1 & -1 & 0 &  1 & -1 & 0 & 0 & 0 & 0\\
0 &  0 & 0 &  0 &  0 & 0 & 0 & 0 & 0\\
1 &  1 & 0 & -1 & -1 & 0 & 0 & 0 & 0\\
1 & -1 & 0 & -1 &  1 & 0 & 0 & 0 & 0\\
\hline
0 &  0 & 0 &  0 &  0 & 0 & 0 & 0 & 0\\
0 &  0 & 0 &  0 &  0 & 0 & 0 & 0 & 0\\
0 &  0 & 0 &  0 &  0 & 0 & 0 & 0 & 0\\
0 &  0 & 0 &  0 &  0 & 0 & 0 & 0 & 0
\end{array}
\right),
\label{eq:Hadamard}
\end{equation}
where the upper-left block corresponds to the logical qubit subspace, and the
remaining entries account for the ancilla degrees of freedom.

In the second stage, a family of high-fidelity control solutions is obtained
for the isolated system.
These unitary reference solutions achieve infidelities well below typical
fault-tolerance thresholds, with
\(
\mathrm{IF} < 10^{-4}
\),
where the infidelity is defined as \(\mathrm{IF} = 1 - F\), and \(F\) denotes the
gate fidelity Eq. (\ref{eq:fidelity}).

In the third stage, the system is coupled to a thermal bath.
The resulting increase in infidelity relative to the unitary reference
solution is analyzed as a function of the relaxation rate and temperature.
Finally, optimal control techniques are applied in the open-system setting to
mitigate the noise, and their effectiveness is quantitatively assessed by
comparing the achieved infidelities to the isolated and uncontrolled cases.

Our analysis is based on two complementary diagnostic tools: (i) fidelity-based measures, comparing the noisy implementation to its isolated unitary reference, and (ii) purity-based measures, which capture the irreversible mixing induced by the bath.  We often use logarithmic ratios such as \(\log_{10}(\mathrm{IF}_{\mathrm{noise}}/\mathrm{IF}_U)\)to highlight regimes where thermal noise dominates over coherent control imperfections. The values \({IF}_{U}\) represent the infidelity of the closed (unitary) reference gate, while \({IF}_{{Noise}}\) indicates the infidelity experienced when executing the same gate with thermal noise.

\subsection{Single qubit gate facilitated through ancilla levels.}
\label{subsec:single-qubit}

We first obtain a set of reference unitary solutions with up to three ancilla levels, where the logical gate is driven purely by indirect coupling: the qubit states do not interact directly, and all transitions proceed via the ancilla. For comparison, we also consider a purely direct-control case with no ancillas, as well as hybrid cases that include both indirect and direct interactions.

At a fixed temperature of \(T = 5\), we explore how the bath induces degradation of gate performance. For this task,  we employ the logarithmic ratio \(\log_{10}({IF}_{{Noise}}/{IF}_{U})\), alongside the final-state dependence of the purity on the relaxation rate \(\gamma\). 
\begin{figure}[H]
\centering    \includegraphics[width=0.65\linewidth, angle=-90]{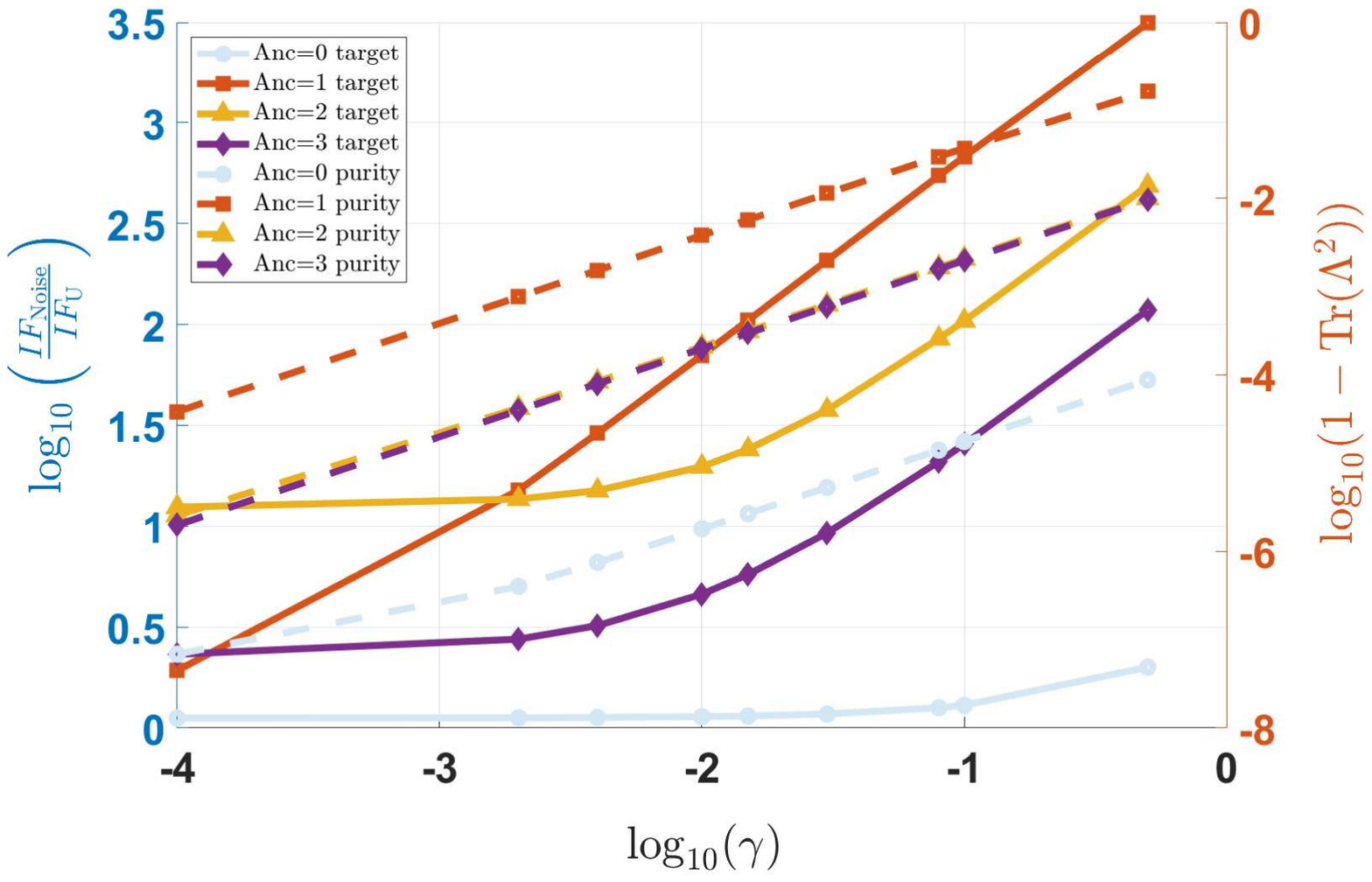}
    
\caption{Normalized infidelity and the purity loss of the map vs the relaxation rate $\gamma $ at fixed temperature \(T = 5\). 
Left axis (solid lines): logarithmic ratio of infidelities 
\(\log_{10}({IF}_{Noise}/{IF}_{U})\) plotted as a function of the system–bath coupling rate \(\gamma\).
Right axis (dashed lines): Generalized purity Eq. (\ref{eq:purity}) of the final reduced qubit map versus \(\gamma\).
 Each color in the graph corresponds to a different number of ancillas coupled to the single qubit. 
} 
    \label{fig:log_ratio_vs_gamma}
\end{figure}
As the relaxation rate \(\gamma\) increases, we can identify three distinct qualitative regimes:

1. {\em Small} $\gamma$: For weak coupling to the bath, the dynamics are close to the closed-system limit, so that $\mathrm{IF}_{\mathrm{Noise}} \approx \mathrm{IF}_{U}$. The log-ratio, therefore, remains close to zero, and the purity is nearly 1. In this range, the curves for different numbers of ancilla levels almost overlap, and the residual differences mainly reflect the dependence of the unitary benchmark $\mathrm{IF}_{U}$ on the system size.

2. {\em Intermediate} $\gamma$: As the bath coupling becomes appreciable, $\mathrm{IF}_{\mathrm{Noise}}$ grows relative to $\mathrm{IF}_{U}$ and the log-ratio becomes positive, accompanied by a noticeable loss of purity. In this region, the curves start to separate, and the influence of the number of ancillas becomes visible. Although the dependence on $N_{\mathrm{anc}}$ is not strictly monotonic at every value of $\gamma$, one can see that, over much of this range, systems with more ancillas are more immune to noise.

3. {\em Large} $\gamma$: For the largest relaxation rates shown, both the log-ratio and the impurity continue to increase in all cases. Over most of this range, the ancilla-assisted cases lie below the curves with fewer ancillas. This indicates that additional ancilla levels better preserve gate fidelity and map purity, even under strong thermal noise.

Taken together, these trends suggest that systematically increasing the number of ancillas helps counteract thermal noise. 

\begin{figure}[H]
    \centering
    \includegraphics[width=0.5\linewidth, angle=-90]{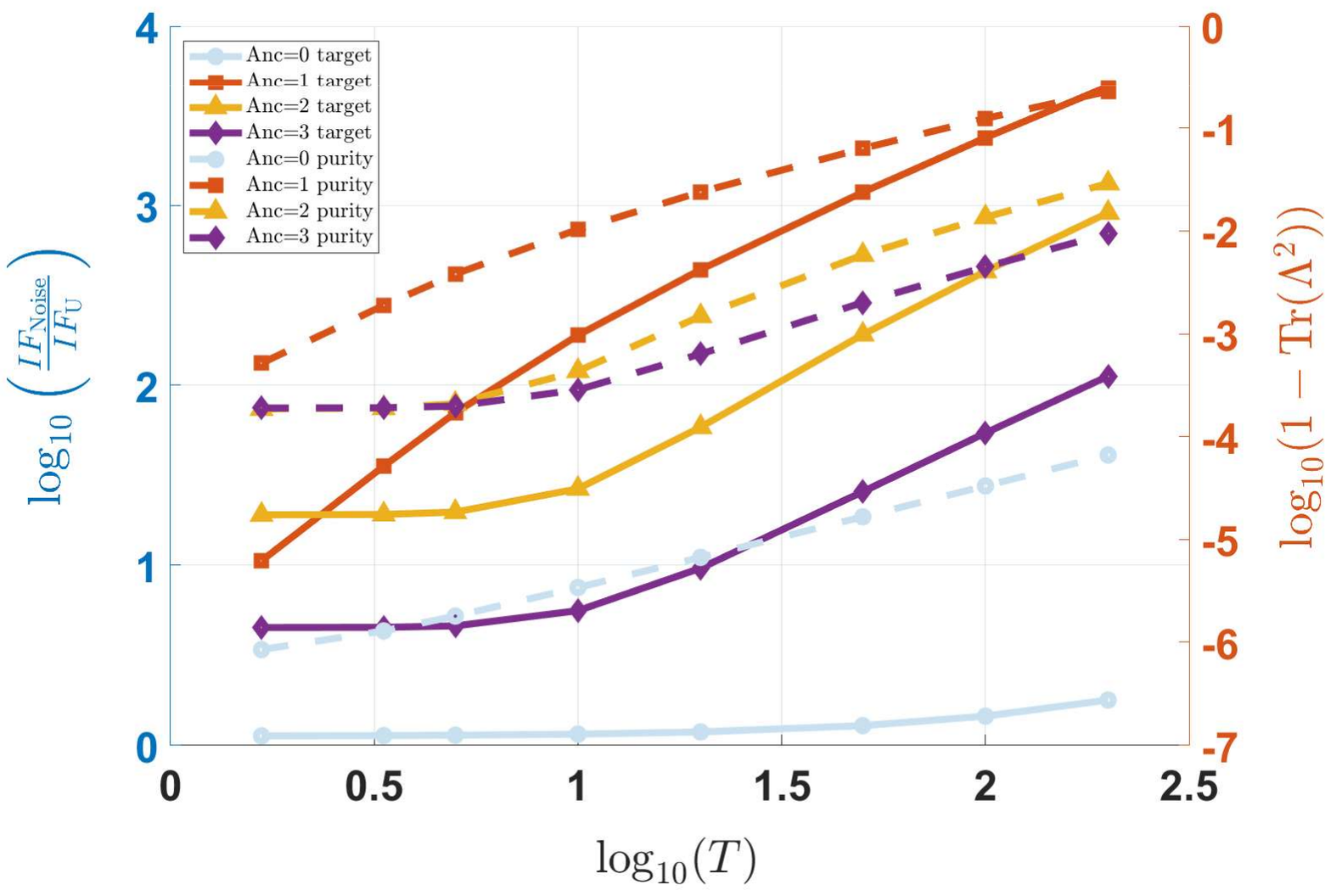}
    \vspace{-1.cm}
    \caption{Normalized infidelity and purity loss of the map with respect to temperature performance at a fixed system-bath coupling rate of \(\gamma = 0.01\).
    Left axis: The logarithmic ratio of infidelities (infidelity loss) \(\log_{10}({IF}_{Noise}/{IF}_{U})\) represented by a solid line against the dimensionless temperature \(T\) (in units of the characteristic transition frequency \(\omega_0\)).The
    Right Axis: the purity of the target map as a function of temperature, indicated by a dashed line. 
    In this context, \({IF} = 1 - {F}\), where \({IF}_U\) is the infidelity of the closed quantum system (as a reference) and \({IF}_{Noise}\) refers to the uncontrolled or naively driven reference state. Each color in the graph corresponds to a different number of ancillas coupled with a single qubit. 
    This representation effectively highlights the effect of temperature on the fidelity and purity of the qubit states.}   
    
    \label{fig:log_ratio_vs_betta}
\end{figure}

Figure \ref{fig:log_ratio_vs_betta} demonstrates the relationship between infidelity ratio and temperature \( T \) at a fixed relaxation rate of \( \gamma = 0.01\). The temperature is expressed in dimensionless units relative to the characteristic transition frequency \( \omega_0 \). 

The log-ratio on the left axis illustrates the benefits of introducing ancillas; positive values indicate that varying thermal noise across temperatures increases the gate's infidelity. Conversely, purity loss, shown on the right axis, increases with rising temperature \( T \), consistent with increased thermal mixing.
At higher effective temperatures, the benefits of additional ancillas appear more pronounced.
\subsection{Logical-map geometry and ancilla participation}
\label{subsec:logical-geometry}

To better analyze the control dynamics,
we need visualization tools that can unravel the complex dynamics of the Hilbert space. For the single-qubit architectures, the logical subspace is spanned by \(\{\ket{0},\ket{1}\}\), while the complementary subspace contains the ancilla levels. We monitor the ancilla population
\begin{equation}
p_A(t)=\mathrm{Tr}\!\left[P_A \rho(t)\right],
\label{eq:pA}
\end{equation}
where \(P_A\) is the projector onto the ancilla sector.

\begin{figure}[H]
    \centering
    \includegraphics[width=0.8\linewidth]{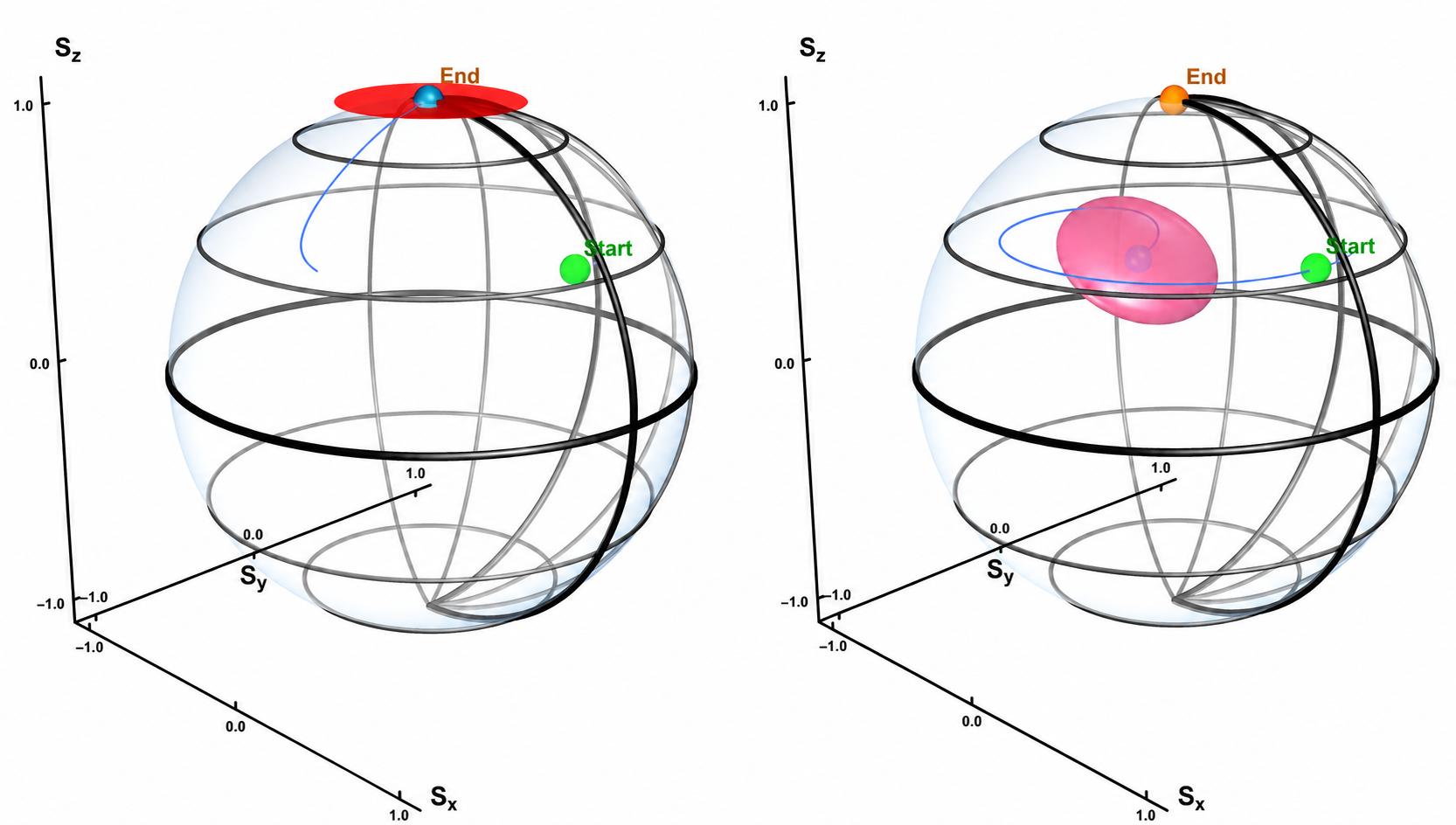}
    \vspace{-.3cm}\caption{
    Visualization of a qubit+ancilla employing Su(3) algebra. The center of the ellipsoid inside the Bloch sphere is determined by the expectation values of
    $\langle S_x \rangle,\langle S_y\rangle ,\langle S_z \rangle$ of the qubit.
    The ellipsoid represents the covariance matrix
    with components $\langle S_i,S_j \rangle$.
    Right: Shown is an intermediate state of the Hadamard gate where the start and end of the transformation are indicated. Left: The discs on the surface of the Bloch sphere represent pure states of the qubit target on the $z$ axis.  (zero population on the ancilla). 
    }
    \label{fig:Su(3)}
\end{figure}

To visualize the complete set of nine operators of the qubit–ancilla system, we employ a generalized Bloch-sphere representation of the SU(3) algebra \cite{h2018non}.
state is depicted as an ellipsoid embedded within the Bloch sphere. Writing the reduced logical state
in an affine form,
\begin{equation}
\vec r \mapsto M(t)\vec r + \vec c(t),
\label{eq:affine-map}
\end{equation}
the singular values of \(M(t)\), denoted \(r_1(t),r_2(t),r_3(t)\), define the principal semiaxes of the Bloch ellipsoid. $\vec c$ is composed from the expectation values of the polarization and defines the center of the ellipsoid.
These quantities characterize the anisotropic contraction of the logical state space induced by the open-system dynamics. Figure~\ref{fig:Su(3)} illustrates a trajectory of the Bloch ellipsoid under the action of the reduced dynamical map. At intermediate times
the trajectory goes through the interior
meaning the ancilla is populated. At the final  time the trajectory reaches the target on the north pole. A disc means zero variance in the $\sigma_z$ direction.
\vspace{-0.3cm}
As a compact measure of the logical-sector deformation, we use the ellipsoid volume
\begin{equation}
\mathcal V_L(t)=r_1(t)r_2(t)r_3(t),
\label{eq:VL}
\end{equation}
which is constant for a volume-preserving unitary map and decreases as the logical dynamics become more dissipative.
\begin{figure}[H]
    \centering    \includegraphics[width=0.6\linewidth,angle=-90]{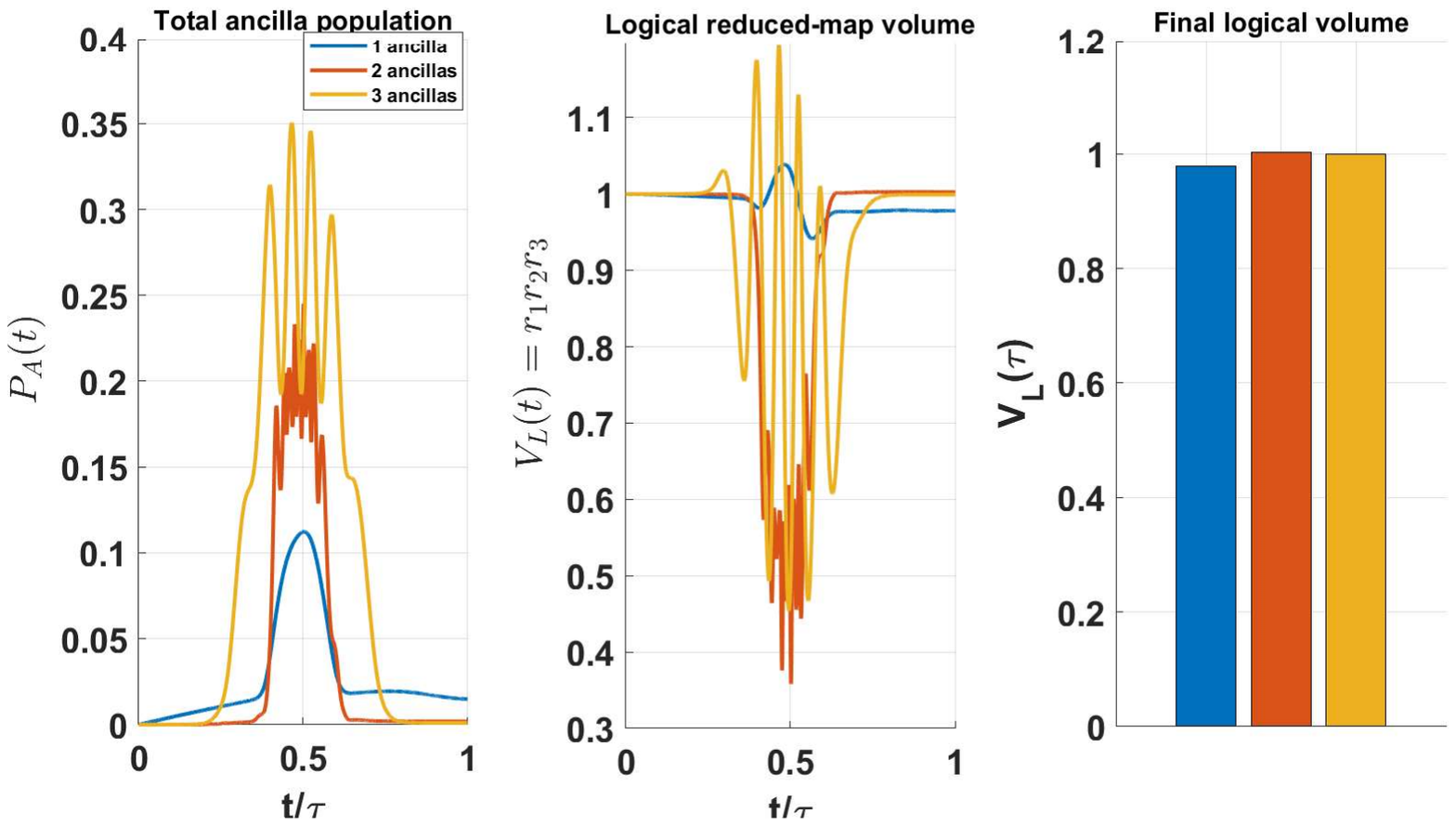}
    \vspace{-1cm}\caption{
    Quantitative characterization of ancilla-assisted mitigation for the single-qubit gate.
    (a) Ancilla population \(p_A(t)\), Eq.~(\ref{eq:pA}), as a function of time for representative optimized protocols.
    (b) Logical Bloch-ellipsoid volume \(\mathcal V_L(t)\), Eq.~(\ref{eq:VL}), extracted from the reduced logical map.
    (c) Corresponding final infidelity (or normalized infidelity ratio) for the same protocols.
    The parameter region in which the ancilla-assisted protocol yields better gate performance is characterized by increased temporary occupation of the ancilla sector and reduced contraction of the logical reduced map.
    This supports, at the level of the present model systems, the interpretation that the optimized dynamics reduces the effective thermal-noise burden on the logical subspace.
    }
    \label{fig:logical_geometry_ancilla}
\end{figure}
Figure~\ref{fig:logical_geometry_ancilla} combines these diagnostics with the corresponding gate-performance measure. The comparison shows that the parameter regime in which ancilla-assisted control improves the gate fidelity is also characterized by two concurrent features: a temporary increase in ancilla occupation and a reduced contraction of the logical Bloch ellipsoid. In other words, the optimized control does not eliminate thermal dissipation globally but instead reshapes the dynamics so that the logical reduced map is less severely degraded.

\begin{figure}
    \centering  \includegraphics[width=0.55\textwidth,angle=-90]{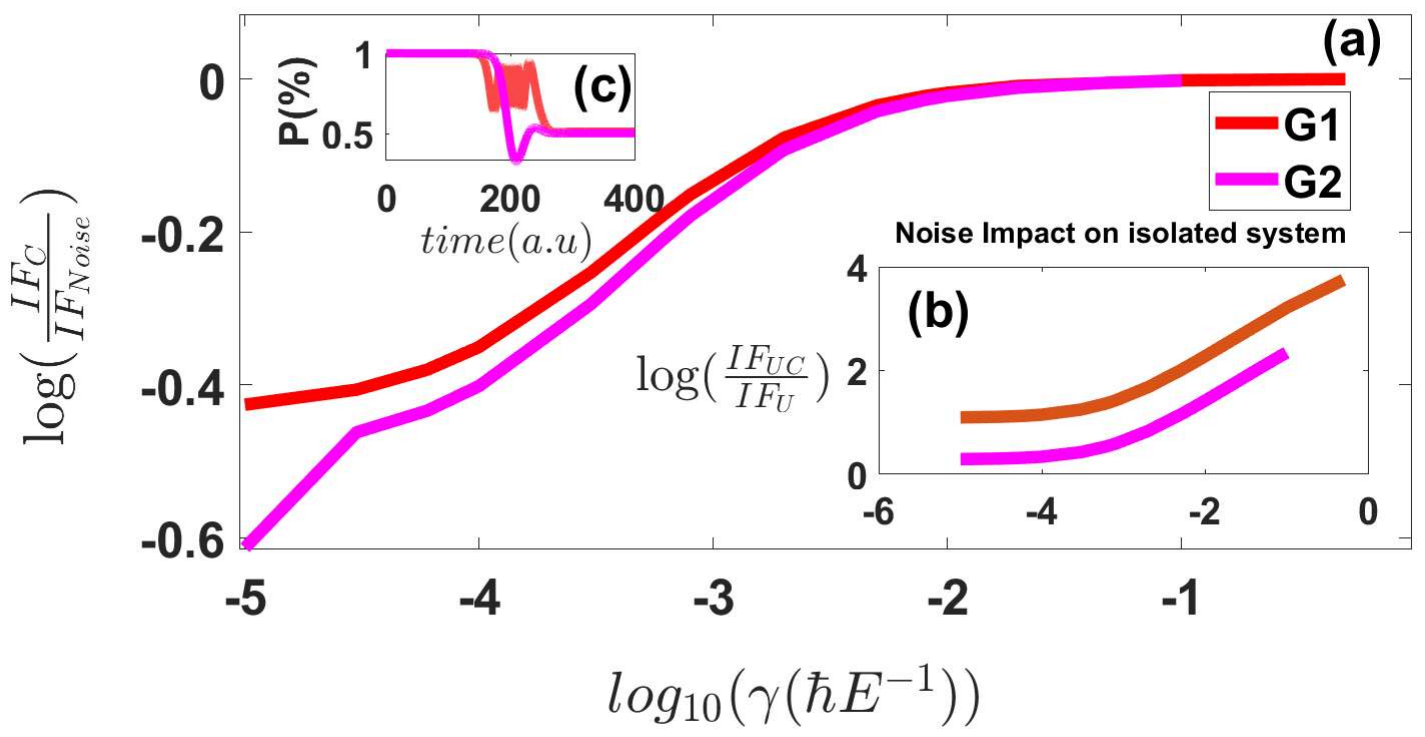}
    
    \caption{Mitigating thermal noise at fixed reduced temperature \(T \approx 10^{-4}\,\omega_{ij}\). (a) Logarithmic of the ratio of the control gain as a function of the relaxation rate $\gamma$: \(G(\gamma)=\log_{10}\!\big({IF}_{C}/{IF}_{Noise}\big)\) vs.\ \(\log_{10}\gamma\). (b) Additional fidelity loss of the unitary reference when coupled to the bath, quantified by $\log_{10}\!\big(\mathrm{IF}_{\mathrm{noise}}/\mathrm{IF}_{U}\big)$; this metric rises monotonically with the relaxation rate $\gamma$. (c) Population projections \(P_i(t)\) (a.u.) under the optimized protocols.Two distinct optimal controls (G1, G2) with different waveform shapes and amplitudes.}
    \label{fig:2_case_Control_1Qbit}
\end{figure}

This interpretation is restricted to the family of models studied here. The analysis does not imply a universal ancilla-protection theorem. Rather, it provides a direct quantitative characterization, within the present multilevel architectures, of how ancilla-assisted control can reduce the effective thermal-noise impact on the logical gate.
Having established the behavior of the gate under noise as a function of both temperature \(T\) (Fig.~\ref{fig:log_ratio_vs_betta}) and system–bath coupling rate \(\gamma\) (Fig.~\ref{fig:log_ratio_vs_gamma}), we have a benchmark to frame the extent by which the environment degrades the gate performance.
In this indirect control of the qubit via ancillas, we attempted to use OCT to mitigate the noise, but achieved only marginal success. The indirect control could not influence the operators
that couple to the environment, and therefore, the OCT had a marginal impact.
\\
\\
\\
To overcome this limitation, we introduce explicit control fields that directly drive transitions between the qubit states. Specifically, an additional $\sigma_x$ interaction is incorporated into the Hamiltonian, providing direct control of the logical qubit. The right inset of Fig.~\ref{fig:2_case_Control_1Qbit} illustrates the degradation of gate fidelity as a function of the relaxation rate \(\gamma\). Results are shown for two representative control protocols. The additional control degree of freedom enables the optimal-control algorithm to substantially suppress the effects of thermal relaxation, leading to improved fidelities over a broad range of dissipation strengths. However, as \(\gamma\) increases, relaxation ultimately dominates the dynamics, and the benefit of optimal control gradually diminishes.

Figure~\ref{fig:2_case_Control_1Qbit} compares two optimized control protocols, \textbf{G1} (red) and \textbf{G2} (magenta), designed for the same system, drift Hamiltonian, and target gate; they differ only in the optimized control waveform \(\varepsilon(t)\). Panel~(a) shows the log–ratio of infidelities \(\log_{10}(\mathrm{IF}_{\mathrm{noise}}/\mathrm{IF}_U)\) versus \(\log_{10}\gamma\). As \(\gamma\) increases, the ratio rises for both protocols, indicating larger fidelity loss with stronger system–bath coupling. The trend exhibits three previously discussed qualitative regimes: a coherent–error–limited regime at very small \(\gamma\), a noise–shaped intermediate regime where control scheduling matters, and an overdamped/Zeno–like regime at large \(\gamma\). Panel~(b) shows the noise effect on the unitary gate: it quantifies how the reference (noise–free) unitary implementation would degrade under the same bath, thereby separating the intrinsic coherent baseline from the thermal contribution. Panel \textbf{(c)} Population projections \(P_i(t)\) (a.u.) under the optimized protocols, highlighting differences between the two control waveforms via distinct population dynamics throughout the gate. To supplement these observation we performed optimization runs employing GRAPE.
The unitary solutions are markedly different and seem to be more sensitive to noise Cf. Appendix \ref{sec:grape}.

\subsection{Landscape traps, initial guesses, and the role of direct qubit control}

Optimizing high-fidelity gates in this \(d=3\) architecture, where one qubit is coupled via an ancilla, is highly sensitive to the landscape of the optimization problem. We observed multiple locally attractive solutions, or {\it good guesses}, along with numerous traps, consistent with reports on non-convex control landscapes in constrained settings. 
\\
In practice, Krotov’s method can still identify successful protocols, but this is effective only when initialized near a basin that contains feasible spectra and timing. When thermal noise affects all system states, trying to combat it solely through ancilla-mediated pathways limits the controller's effectiveness. The drive must not only implement the desired unitary operation but also avoid frequencies \(\omega_{ij}(t)\) where \(J(\omega)n_T(\omega)\) is large. This requirement is not always compatible with coupling access that relies solely on the ancilla.

This limitation is particularly evident at (T=5) in Fig.~\ref{fig:log_ratio_vs_gamma}. In this regime, the control protocol can implement only a limited number of corrective actions, or the required corrections become prohibitively costly over a broad range of relaxation rates \(\gamma\). Consequently, the improvement provided by optimal control is marginal, yielding only modest gains in fidelity.

However, introducing even a weak direct controlled coupling between the qubit states, specifically a second controller addressing the transition \(|0\rangle\!\leftrightarrow\!|1\rangle\), significantly enhances robustness. This improvement is illustrated in Fig.~\ref{fig:2_case_Control_1Qbit}, where the mitigation increases the fidelity by an order of magnitude.

Operationally, we exploited this by first optimizing the control field in a model without direct coupling, and then applying the resulting field to a slightly perturbed system that includes a weak direct interaction of order \(\sim 10^{-3}\).
Relative to the indirect coupling strength. This weak perturbation preserves the qualitative structure of the previously obtained solution, while introducing just enough modification to the control landscape topology for Krotov’s method to further refine the field toward a higher-fidelity unitary in the closed (noise-free) model. The resulting optimized field was then used as the initial guess for the thermal (open-system) dynamics, thereby reducing the gate error in the presence of dissipation. 

In contrast, when the direct qubit interaction is increased beyond the weak-perturbation regime, the effectiveness of the transferred solution deteriorates. The original field is then driven away from a favorable basin of attraction, and the subsequent optimization behaves effectively as a reinitialization from a random initial guess. This illustrates a central practical lesson in optimal control: {\em high-quality initial guesses are essential}, particularly in quantum gate design. More broadly, it highlights that small, structured perturbations can transform otherwise intractable searches in complex control landscapes into manageable local refinements.

\subsection{Two-qubit C-iX gate under thermal noise}
\label{sec:2q-CiX}

A universal set of one and two-qubit gates requires
at least one entangling gate
\cite{muller2011optimizing}.
The choice of the two-qubit gate was motivated by our previous study on mitigating controller noise \cite{aroch2024mitigating}. In that work, the dominant noise mechanism was dephasing, modeled by a double-commutator structure
${\cal L}_D \propto [ \hat H_m, [\hat H_m,\bullet]] $ where $\hat H_m$ is either the total Hamiltonian or its time dependent part $\hat H_t$. 
Thermal noise differs in that the environment can exchange energy with the system. This enables a generalized cooling mechanism that can actively reduce noise \cite{kallush2022controlling}. In the high-temperature limit, the thermal dissipator approaches a pure dephasing form, reducing to the familiar double-commutator structure \cite{pyurbeeva2025bloch}.

The logical two-qubit subspace consists of $\{\ket{00}, \ket{01}, \ket{10}, \ket{11}\}$.
The full Hilbert-space dimension is therefore $d = 4 + N$.
In this enlarged space, we define the drift Hamiltonian
\begin{equation}
    \hat{H}_0 = \omega\, {G}_{\text{qubit}} + \sum_{j=1}^{N} \Delta_j\, {G}^{(a_j)}, 
    \qquad \Delta_j = 4j \omega,
    \label{eq:H0-2q-ancilla}
\end{equation}
with
\[
    {G}_{\text{qubit}} = \mathrm{diag}(-1, 0, 0, 1, 0, \dots, 0).
\]
\\
Interactions between ancillas and specific two-qubit basis states $\alpha \in \{00, 01, 10, 11\}$ are written as
\begin{align}
    G^{(j)}_\alpha &= \ket{\alpha}\!\bra{a_j} + \ket{a_j}\!\bra{\alpha}, \\
    \hat{H}_{c1}(t) &= \sum_{j=1}^{N} \sum_{\alpha} \epsilon^{c1}_{\alpha,j}(t)\, G^{(j)}_\alpha, \\
    \hat{H}_{c2}(t) &= \sum_{j=1}^{N} \sum_{\alpha} \epsilon^{c2}_{\alpha,j}(t)\, G^{(j)}_\alpha ,
    \label{eq:Hc-ancilla-2q}
\end{align}
where the controlled amplitudes $\epsilon^{c1}_{\alpha,j}(t)$ and $\epsilon^{c2}_{\alpha,j}(t)$ generate the desired logical coupling.

In this work, we focus on the thermal-noise study of the direct-control realization without ancillas ($N=0$), where $d=4$ and the target gate is the controlled-$iX$ (C-iX) gate.
The C-iX operation acts as the identity when the control qubit is in $\ket{0}$ and applies $i \hat{X}$ on the target qubit when the control is in $\ket{1}$,
\begin{equation}
    \hat{U}_{\text{C-iX}}
    = \ket{0}\!\bra{0} \otimes \hat{\mathcal{I}}
      + \ket{1}\!\bra{1} \otimes i\hat{X},
    \label{eq:CiX-def}
\end{equation}
so that this entangling gate is suitable as a nontrivial benchmark for open-system control.

For the direct two-qubit implementation, we employ the uncorrelated drift Hamiltonian
\begin{equation}
    {\hat H}_0= 
    a \tilde {\cal{I}}^1\otimes \tilde{\cal{I}}^2 +\omega_1 
    {\tilde \sigma}_Z^1\otimes \tilde{\cal{I}}^2,
    \label{free-dynamics-hadamard3}
\end{equation}
where $a$ is a phase factor, and $\omega_1$ is the qubit frequency.
Increasing the number of control fields is essential for creating this gate in a two-qubit space:
\begin{equation}
    {\hat H}_{c1}= \varepsilon_1(t)\sum_{i=X,Y}a_i(\tilde {\cal I}-{\tilde \sigma}_Z)^1 \otimes {\tilde \sigma}_i^2 ,
    \label{control-dynamics-CNOt_E_1}   
\end{equation}
\begin{equation}
    { \hat H}_{c2}= \varepsilon_2(t)({\tilde \sigma}_X^1 \otimes {\tilde \sigma}_Z^2),
    \label{control-dynamics-CNOt_E_2}
\end{equation}
where the field \({ \hat H}_{c2}\) introduces a correlation between the qubits.
The Hamiltonian can also be expressed in terms of Gell-Mann matrices; for clarity, it is written here as a sum of Pauli matrices.
In this expression, \(\tilde{\sigma}^{(k)}_i\) denotes the Pauli generators (\(i=X, Y, Z\)) for qubit \(k\) in the Liouville representation, with \(\varepsilon_{1/2}(t)\) acting as the control envelope and \(a_i\) representing fixed real coefficients.
This unified control mechanism will be used to generate our target gate, a two-qubit entangling gate \(\hat{U}_{\text{C-iX}}\),
\begin{align}
\hat U_{\text{C-iX}}=
\left(\begin{array}{llll}
1 & 0 & 0 & 0 \\
0 & 1 & 0 & 0 \\
0 & 0 & 0 & i \\
0 & 0 & i & 0
\end{array}\right)
\end{align}
 in an isolated scenario.
To investigate the degradation and mitigation of the C-iX gate under thermal GKLS noise, we employ the Liouvillian framework where \(\boldsymbol{\Lambda}(\tau)\) denotes the propagated map.
Thermal noise is represented by a dissipative Liouvillian \(\boldsymbol{\mathcal L_D}\) that links the qubits to a bosonic bath at temperature \(T\), adhering to detailed balance for both upward and downward rates. 

For each combination of \((T,\Gamma)\) values, with \(\Gamma\) being the overall thermal rate scale, the Krotov algorithm is utilized to optimize the control fields \(\varepsilon_{1,2}(t)\).
The objective is to ensure that the implemented map on the subspace \(\boldsymbol{\Lambda}_{\mathrm{sub}}(\tau)\) closely approximates the ideal C-iX map within the logical subspace.
Performance evaluation involves comparing the noisy two-qubit gate to its unitary reference and monitoring metrics such as gate fidelity and Liouville-space properties, including subsystem purity, which is relevant to \(\boldsymbol{\Lambda}_{\mathrm{sub}}(\tau) \).

We first quantify the infidelity loss by comparing the noisy map $\boldsymbol{\Lambda}_{\mathrm{sub}}(\tau)$ with its unitary reference $\boldsymbol{\Lambda}_U(\tau)$.
Denoting by $\mathrm{IF}_U$ the infidelity of the isolated C-iX and by $\mathrm{IF}_\mathrm{noise}$ the infidelity in the presence of the thermal bath, we define
\[
    R_\mathrm{IF} = \frac{\mathrm{IF}_\mathrm{noise}} {\mathrm{IF}_U} ,
\]
which measures degradation solely due to environmental interactions.
\begin{figure}[H]
    \centering  \includegraphics[width=0.47\textwidth,angle=-90]{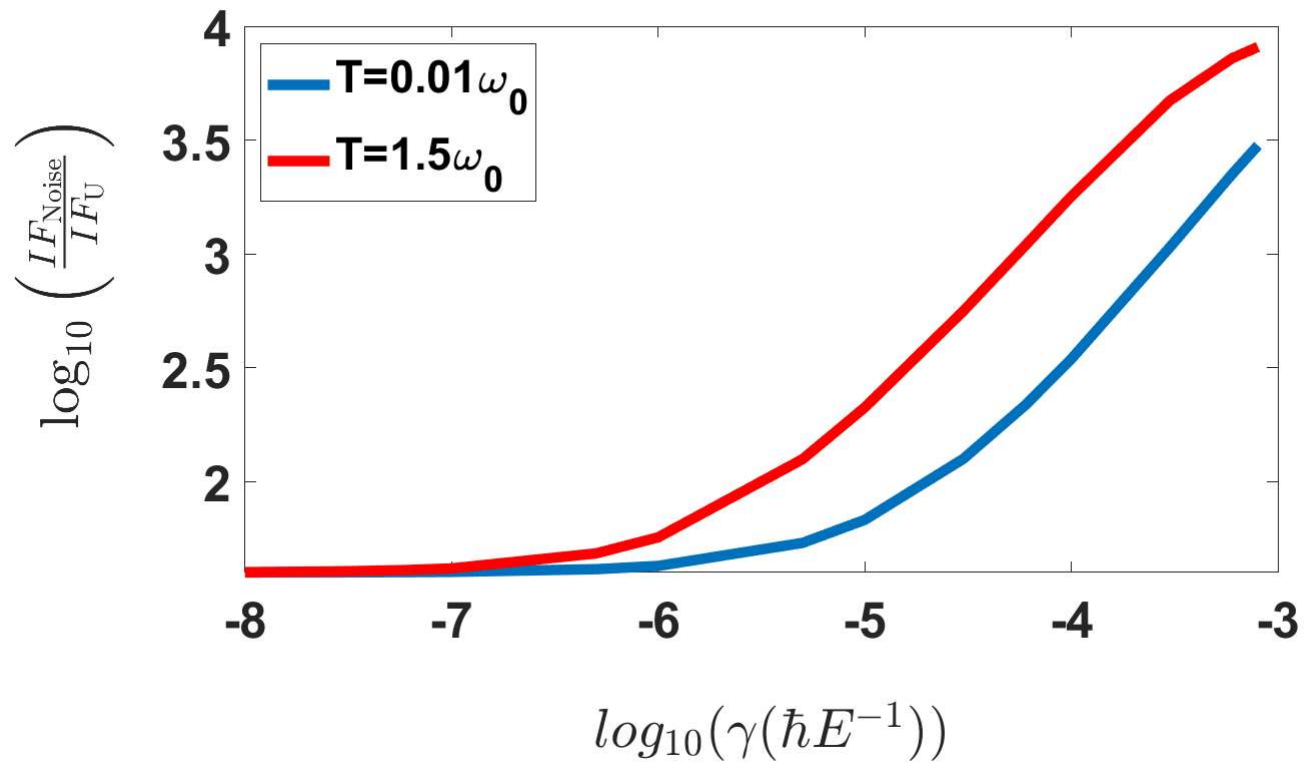}
    \caption{Normalized infidelity $R_{IF}$ as a function of relaxation rate $\gamma$ for the two-qubit C-iX gate. Hot and cold temperatures $T$ are shown.
    Here, $\mathrm{IF}_U$ denotes the infidelity of the ideal isolated C-iX reference, while $\mathrm{IF}_\mathrm{noise}$ is the infidelity obtained in the presence of the thermal GKLS dynamics.
    The plot highlights a low-noise region where $R_{\mathrm{IF}}$ remains small and the optimized control essentially reproduces the unitary benchmark, and a high-noise region where the environmental contribution dominates the gate error.}
    \label{fig:2Q-IF-loss}
\end{figure}
Figure~\ref{fig:2Q-IF-loss} summarizes the extent of additional error generated by the thermal bath across the $(\gamma, T)$ plane. For small $\gamma$ and low $T$, the infidelity loss $R_{\mathrm{IF}}$ is close to zero, indicating that the optimized C-iX pulse is robust, and the residual error budget is essentially the same as in the isolated case. As either $\gamma$ or $T$ is increased, $R_{\mathrm{IF}}$ grows, reflecting the growing influence of thermally activated transitions on the gate. The plot thus clearly separates a {\em control-dominated regime}, where coherent imperfections are the main limitation, from a {\em noise-dominated regime} in which thermal processes set the ultimate accuracy.
As either parameter is increased, the ratio degrades, demonstrating that thermally induced transitions increasingly compete with the coherent dynamics.
The fan-out of the curves with $T$ reflects the expected trend: hotter baths accelerate the loss of performance at a given coupling strength.

\begin{figure}[H] 
    \centering
    \includegraphics[width=0.47\textwidth,angle=-90]{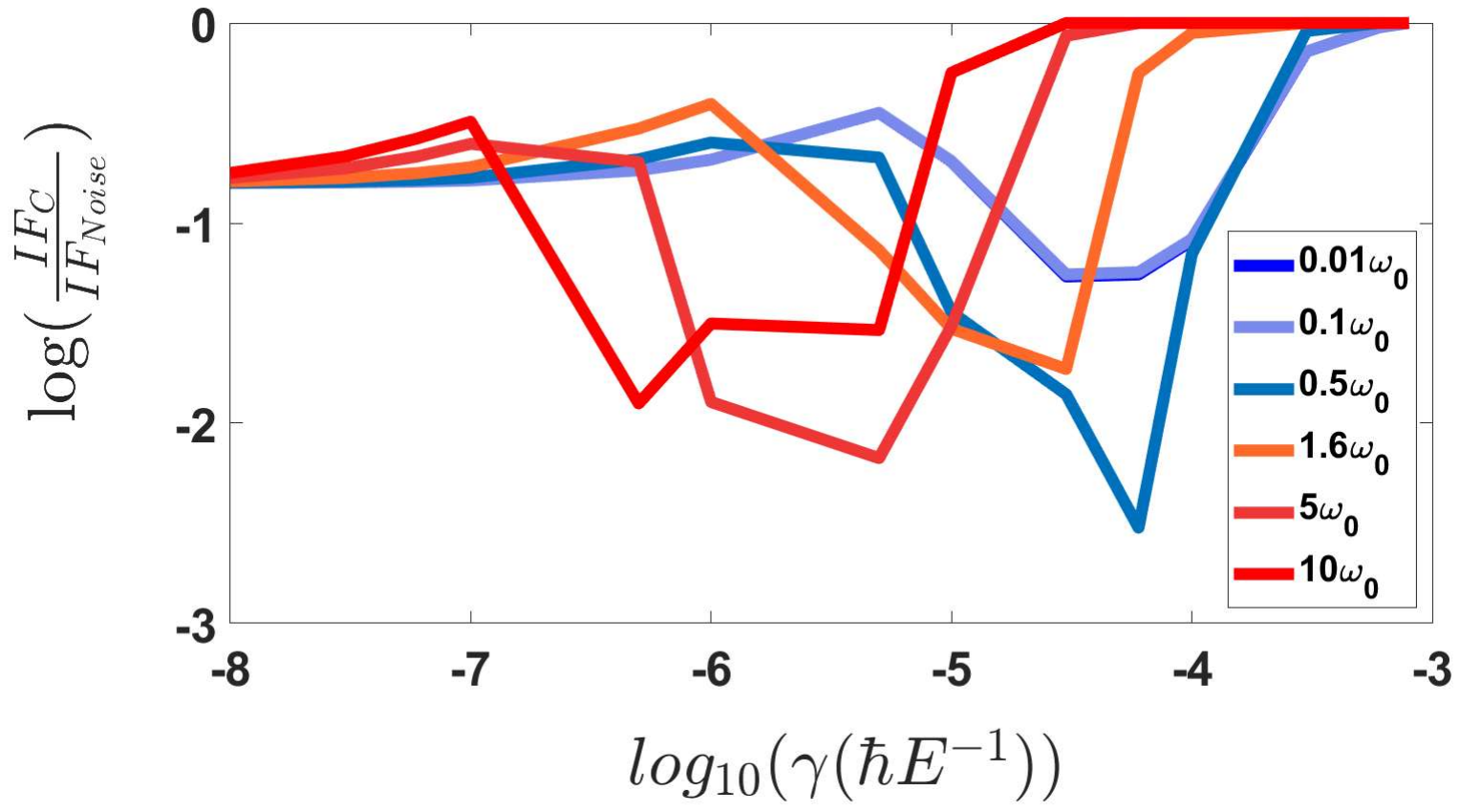}
    \caption{ Mitigation gain for the two-qubit C-iX gate as a function of the thermal rate $\gamma$ at several bath temperatures $T$.
    The gain is defined as the ratio of the infidelity obtained with OCT in the presence of noise to the infidelity without OCT corrections.
    The curves for $T=0.01\,\omega_0$ and $T=0.1\,\omega_0$ nearly coincide and follow the same trajectory; we revisit their peak locations in Fig.~\ref{fig:2Q-purity-ss}.}
    \label{fig:2Q-infidelity-ratio}
\end{figure}

The next step is to check if Optimal Control Theory (OCT) with direct controllers can mitigate this thermal noise. Figure~\ref{fig:2Q-infidelity-ratio} shows 
the mitigation gain of the optimized C-iX gate as a function of the thermal rate $\gamma$ for several bath temperatures $T$ (different colors). The temperature is expressed in dimensionless units relative to the characteristic transition frequency \( \omega_0 \). 
Examining Fig.~\ref{fig:2Q-infidelity-ratio}, we observe that at low temperatures and low thermal rates, the infidelity ratio remains close to unity. This indicates that the optimized pulse closely reproduces the isolated C-iX gate, with only a minor additional error induced by the bath. Consequently, the degree of error correction achieved by OCT in this regime is modest. As the temperature increases, a distinct optimal window emerges in which error mitigation becomes highly effective, suppressing errors by up to two orders of magnitude. However, at sufficiently large 
$\gamma$ values, OCT can no longer compensate for the noise, and the mitigation window closes. It is also evident that error reduction is considerably more effective at lower temperatures.

\begin{figure}[H]
  \centering \includegraphics[width=0.47\textwidth, angle=270]{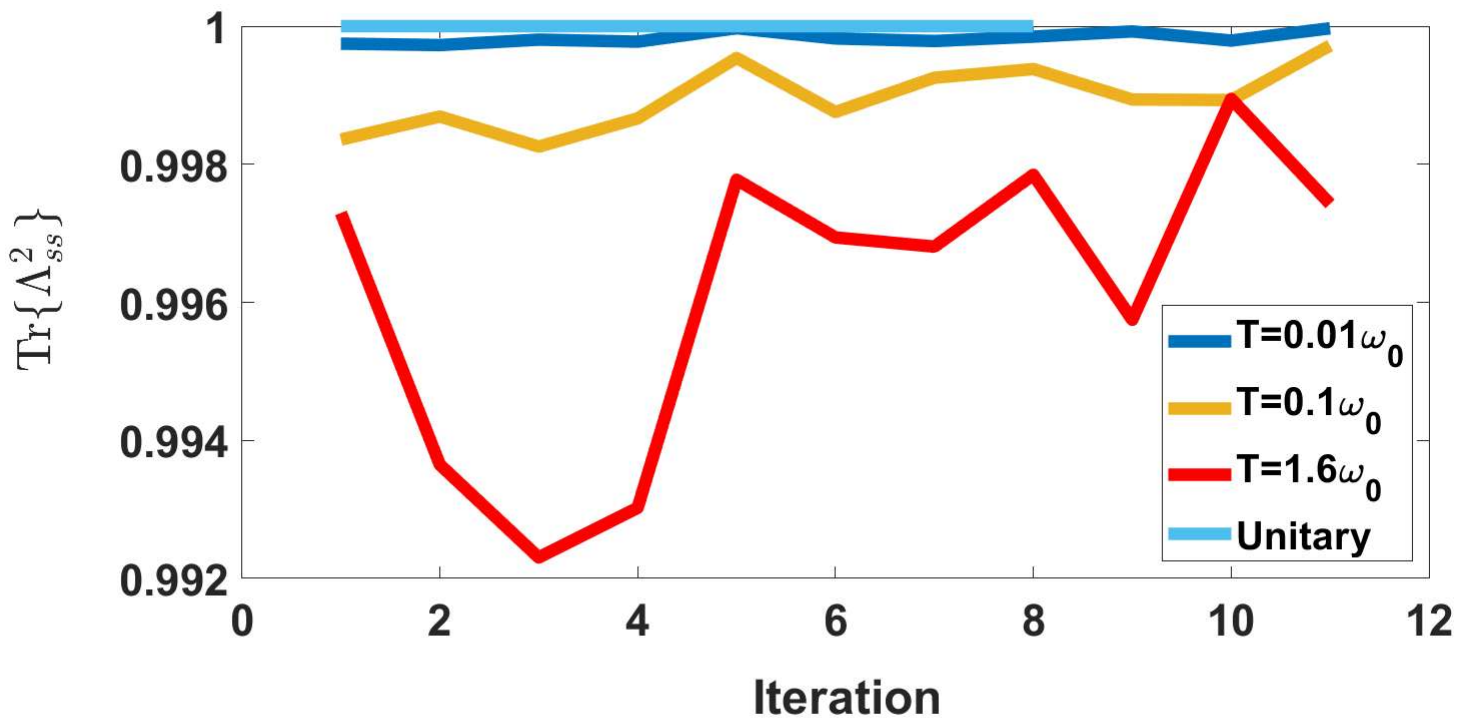}
  \caption{Subspace channel purity \(\mathcal{P}_\mathrm{sub}
  = \mathrm{Tr}\!\{\boldsymbol{\Lambda}_\mathrm{sub}^\dagger
  \boldsymbol{\Lambda}_\mathrm{sub}\}/M^{2}\) for the two-qubit C-iX gate
  as a function of the OCT iteration index at fixed \(\gamma=3\times10^{-5}\),
  for several temperatures \(T\). The map purity is computed from the restricted map \(\boldsymbol{\Lambda}_\mathrm{sub}\) on the working set of operator directions. The purity remains high in all cases and decreases relatively slowly as \(T\) increases.}
  \label{fig:2Q-purity-ss}
\end{figure}

To connect with the Liouville-space mechanism that will be discussed in Sec.~\ref{sec:discussion}, we reconstruct the map $\boldsymbol{\Lambda}(\tau)$ on a complete operator basis of size $N^2 = 16$ and then restrict it to the subset of operator directions on which the C-iX gate acts. 
This defines a reduced map $\boldsymbol{\Lambda}_{\mathrm{sub}}$ and a corresponding subsystem purity of the map (Choi state purity) \cite{de2026dynamical,gour2021entropy}, 
\begin{equation}
    {\cal P}_\mathrm{sub} \equiv \frac{1}{M} \mathrm{Tr}\bigl(\boldsymbol{\Lambda}_\mathrm{sub}^\dagger \boldsymbol{\Lambda}_\mathrm{sub}\bigr).
    \label{eq:purity}
\end{equation}
where $M$ is the dimension of the subsystem.
We examine the behavior of the subspace purity in a regime where optimal control
mitigates the error by one to two orders of magnitude (see Fig.~\ref{fig:2Q-infidelity-ratio}). Figure~\ref{fig:2Q-purity-ss} shows \(\mathcal{P}_{\mathrm{sub}}\) versus the OCT iteration index for several temperatures \(T\) at fixed \(\gamma=3\times10^{-5}\). In all cases, the purity remains high, with a modest improvement from the first to the last iteration. At higher \(T\), the optimization \emph{uses purity as a resource}: early iterations may transiently lower \(\mathcal{P}_{\mathrm{sub}}\) to gain fidelity, followed by a recovery phase in which purity is restored within the working subspace as the gate converges. The unitary (no-bath) baseline remains near
\(\mathcal{P}_{\mathrm{sub}}\!\approx\!1\) throughout.

Examining Fig.~\ref{fig:2Q-purity-ss}, the subspace purity 
\(\mathcal{P}_\mathrm{sub}\) stays close to its maximal value for weak noise, indicating that the logically relevant part of the map is nearly unitary. As \(T\) increases, \(\mathcal{P}_\mathrm{sub}\) is gradually reduced but remains significantly high. This behavior is consistent with trajectories that move slightly within the Bloch sphere while also undergoing rotation, both of which contribute to infidelity.

\begin{figure}
    \centering   \vspace{-2.5cm}\includegraphics[width=0.8\textwidth,angle=-90]{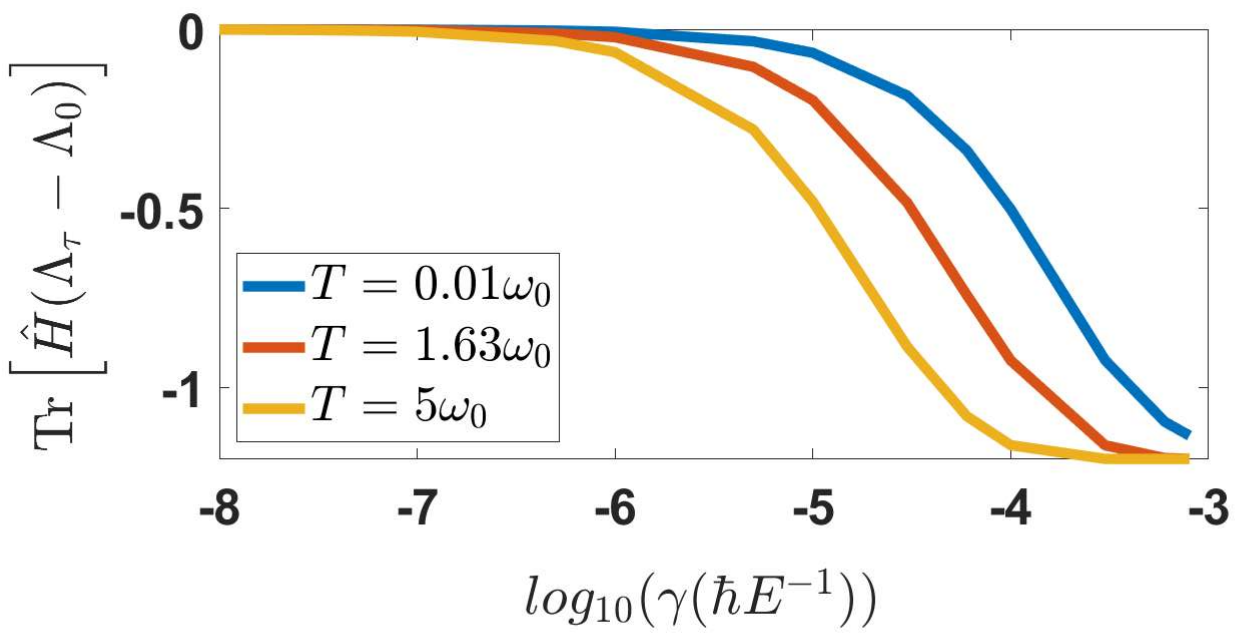}
    \vspace{-3cm}\caption{The change in energy  $\Delta E$ 
    as a function of the relaxation rate $\gamma$ for different bath temperatures during the two-qubit C-iX gate.
    The quantity $\Delta E$ is computed from the evolution of the drift Hamiltonian under the full map $\boldsymbol{\Lambda}(\tau)$ Eq. (\ref{eq:echange}).
    Negative values correspond to a net energy flow from the system to the bath, while regions with small $|\Delta E|$ indicate nearly energy-conserving operation.} 
    \label{fig:2Q-energy-loss}
\end{figure}

Beyond this structural information in Liouville space, it is helpful to look at the net energy exchanged with the bath during the gate.
We next quantify the energetic signature of the thermal bath by monitoring the action of the map on the drift Hamiltonian in Liouville space at the finite time $\tau$,
\[
    \vert H_0(\tau)\rangle\rangle = \boldsymbol{\Lambda}(\tau)\,\vert H_0(0)\rangle\rangle.
\]
The corresponding energy change imposed by the gate becomes,
\begin{equation}
    \Delta E = \mathrm{tr}\{\hat{H}_0 (\tau)\} - \mathrm{tr}\{\hat{H}_0(0) \},
    \label{eq:echange}
\end{equation}
Eq. (\ref{eq:echange}) is equivalent to
$\Delta E = \mathrm{tr}\{\hat{H}_0 (\tau) \hat \rho \} - \mathrm{tr}\{\hat{H}_0(0)\hat \rho \},$
where $\hat \rho \propto \hat I$ is at infinite temperature. As a result, this gate energy measure is invariant to any unitary
transformation. Therefore, any change in energy is due to energy transfer to the environment.
The result is shown in Fig.~\ref{fig:2Q-energy-loss} as a function of $\gamma$ and $T$.

Fig.~\ref{fig:2Q-energy-loss} reveals how the C-iX implementation exchanges energy with the thermal environment.
For weak coupling and low temperature, the energy change is small, consistent with the gate operating close to an isolated, quasi-unitary trajectory.
As $\gamma$ and/or $T$ increase, $\Delta E$ becomes more negative, showing that the bath increasingly acts as an energy sink that removes excitations generated during the driven dynamics. The irreversibility of the 
process can be characterized by the entropy production in the bath $\Delta {\cal S}_u =-\frac{\Delta E}{T}$. Examining Fig. \ref{fig:2Q-energy-loss}, the entropy generation of the cold bath is $\sim 50$ larger than that of the high-temperature environment.
Considering also Fig.~\ref{fig:2Q-IF-loss}, this confirms the dissipation-assisted control picture: in the low-noise regime, the optimized C-iX behaves almost energetically neutral and close to its unitary reference, whereas in the strong-noise regime, the bath both degrades the fidelity and drains energy from the system during the gate.

We now examine combined error sources by adding a
controller phase noise with the form:  \cite{aroch2024mitigating}: $${\cal D}_P =  -{\gamma}_P[\hat{{H}_S}(t),[\hat{H}_S(t),\bullet]]~~.$$ For example, 
for a thermal noise of $\gamma=4. 10^{-5} $
and temperature of $T=0.5 $ we added phase noise of $\gamma_P=10^{-4}$.
The initial infidelity for thermal noise was $IF=10^{-3}$ and for phase noise $IF=10^{-2}$
combined $IF = 10^{-1}$. In each case, optimal control was able to restore
to at least $IF \sim 10^{-4}$. Combined
optimal control staggered and is unable to restore the fidelity, meaning there is a conflict in mitigating the combined noise.

\section{Discussion}
\label{sec:discussion}

A theoretical analysis can explain the loss of gate fidelity due to thermal noise.
Two distinct mechanisms contribute to the deviation from the target transformation.
The first is a misguided unitary evolution without loss of purity: on the generalized Bloch sphere, the state remains on the surface but ends up in an erroneous final state.
Additional coherent control fields can correct such purely coherent errors.
The second mechanism is an irrecoverable loss of purity generated by the dissipative dynamics:
On the Bloch sphere, this corresponds to motion towards the sphere's interior and reflects entropy production due to coupling to a thermal bath.
\\
For a quantum system evolving under a GKLS generator 
$\boldsymbol{\mathcal L} = \boldsymbol{\mathcal L}_H + \boldsymbol{\mathcal D}$,
The instantaneous generalized purity \cite{de2026dynamical} loss is given by
\begin{equation}
\frac{d}{dt}{\cal P}=
\frac{d}{dt} \textrm{Tr} \{ \hat{\boldsymbol{\Lambda}}^2 \}
~=~2 \textrm{Tr} \{\hat{\boldsymbol{\Lambda}}  \boldsymbol{ \mathcal L} \hat{\boldsymbol{\Lambda}}\} 
~=~2\textrm{Tr} \{\hat{\boldsymbol{\Lambda}}  \boldsymbol{ \mathcal D} \hat{\boldsymbol{\Lambda}}\} ~~,
  \label{eq:purityloss-thermal}
\end{equation}
where the Hamiltonian part $\boldsymbol{\mathcal L}_H[\hat{\boldsymbol{\bullet}}] = -\frac{\mathrm{i}}{\hbar}[\hat{H}(t),{\bullet}]$ conserves purity and only the dissipative part 
$\boldsymbol{ \mathcal D}$ contributes to purity loss.

In the present setting, the gate is implemented over a finite time $T$. 
The interplay between the gate duration and the thermal rates $\gamma(t)$ determines the balance between coherent and incoherent error.
Short control times reduce the total exposure to $\boldsymbol{\mathcal D}$ but require stronger, more rapidly varying fields; longer durations allow the control to follow smoother control solutions but lead to greater thermal relaxation.
Our optimizations show that, for fixed temperature and rates, the residual infidelity cannot be eliminated.
Even after re-optimization, a finite loss of purity remains, reflecting irreversible excitation exchange with the bath.
This behavior is visible in the two-qubit C-iX simulations, where the infidelity ratio and the infidelity loss both increase as $\gamma$ and $T$ grow (Figs.~\ref{fig:2Q-infidelity-ratio} and~\ref{fig:2Q-IF-loss}). At the same time, the subspace purity and the energy change confirm the presence of an irreducible dissipative component (Figs.~\ref{fig:2Q-purity-ss} and~\ref{fig:2Q-energy-loss}).

The control topology also plays a key role.
In the indirect-control architecture (single qubit plus ancilla), enlarging the Hilbert space increases the number of thermally active transitions.
The ancilla levels open additional decay and excitation channels and thus increase the effective exposure to $\boldsymbol{\mathcal L_D}$.
As a result, convergence basins in the control landscape become narrower under thermal noise, and the residual infidelity remains above the unitary reference even after re-optimization.
Allowing a small direct drive on the logical transition suppresses the dominant thermal pathways that directly involve the computational levels, essentially closing this gap.
Increasing the number of ancillas can, on the one hand, reduce effective thermal sensitivity by creating interference paths that bypass the most strongly coupled transitions; on the other hand, in the isolated limit, the search for high-fidelity solutions becomes more demanding due to a more rugged landscape in the enlarged control space.
For the direct-control, two-qubit implementation of the controlled-$iX$ (C-iX) gate, the $(\gamma, T)$ scans clearly separate a control-dominated region from a noise-dominated region:
at low values of $\gamma$ and $T$, the infidelity loss $R_{\mathrm{IF}}$ remains small. The gate closely tracks its unitary reference, whereas at larger $\gamma$ and $T$, thermal processes dominate and the fidelity decreases monotonically (cf.\ Sec.~\ref{sec:2q-CiX}, Figs.~\ref{fig:2Q-infidelity-ratio} and~\ref{fig:2Q-IF-loss}).

\medskip

Beyond the state-based picture of Eq.~(\ref{eq:purityloss-thermal}), additional insight into the control mechanism is obtained by analyzing the dynamics in Liouville space.
Instead of following a single density operator, we propagate a complete operator basis of dimension $N^2$ and reconstruct the map $\boldsymbol{\Lambda}(\tau)$ generated by the thermal GKLS evolution at the final time $T$.
For the two-qubit C-iX gate, the relevant logical dynamics are embedded in this larger space but effectively act on a smaller subset of basis operators (about ten in our case) on which the optimal control performs non-trivial work.

Restricting $\boldsymbol{\Lambda}(T)$ to this ``working'' subspace defines a reduced map $\boldsymbol{\Lambda}_\mathrm{sub}$, whose Hilbert--Schmidt norm, Eq. (\ref{eq:purity}). 
${\cal P}_\mathrm{sub}$ is interpreted as the subsystem purity of the gate map on the computational subspace.
We find that optimal control increases this quantity compared to the uncontrolled thermal dynamics:
Within the working subspace, the map becomes more unitary-like, even though the global evolution governed by $\boldsymbol{\mathcal D}$ remains non-unitary.
This is reflected in the relatively slow decay of $P_\mathrm{sub}$ with $\Gamma$ and $T$ in Fig.~\ref{fig:2Q-purity-ss}.
By contrast, the complementary basis operators, which are not addressed by the gate functional, remain almost invariant under the dynamics and retain purities very close to unity, $\sim(1-10^{-12})$.
The control thus sculpts an effectively decoherence-resilient subspace in Liouville space: the relevant operator directions are rotated into combinations of eigen-operators of $\boldsymbol{\mathcal L}$ that are only weakly damped, while dissipation acts predominantly on directions that are irrelevant for the logical gate.

A second diagnosis comes from the action of the map on the system Hamiltonian in Liouville space.
Writing $\vert H_0(\tau)\rangle\rangle = \boldsymbol{\Lambda}(\tau)\,\vert H_0(0)\rangle\rangle$, we observe a net  negative energy change between the initial and final times,
$
    \Delta E = \mathrm{tr}\{ \hat{\Lambda}(\tau) \hat{H}_0\} - \mathrm{tr}\{ \hat{\Lambda}(0) \hat{H}_0\},
$
as summarized in Fig.~\ref{fig:2Q-energy-loss}.
This indicates that, in parallel to the coherence protection on the logical subspace, the environment acts as an energy sink: thermal relaxation removes excitations, and the control field steers the system along trajectories where this energy change is compatible with maintaining high gate fidelity.
The combined picture is therefore one of dissipation-assisted control:
The optimal field not only suppresses thermal GKLS noise but also reshapes the effective Liouvillian seen by the computational subspace, trading global energy relaxation for increased subsystem purity in the map and the fidelity of the implemented C-iX gate.
Hints for this mechanism have been revealed 
in \cite{kallush2022controlling} for gates in a thermal environment. That study employed an approximate master equation with fixed temperature.
Finally, when we add phase noise to the thermal noise, we find that the optimal control has difficulty in mitigating both noise sources simultaneously. The two correction protocols contradict each other; consequently, the control cannot significantly improve gate fidelity.

\section{Conclusions}
\label{sec:conclus}

Quantum devices employ interference and entanglement as crucial resources, while dissipation remains a primary limiting factor \cite{haffner2008quantu,preskill2018quantum,schlosshauer2019quantum,kosloff2024quantum}. 
When engineering a quantum gate, faithful dynamical simulation is a crucial tool. A central aspect of this study is the incorporation of thermal noise into the model. The main challenge lies in determining the control field, which simultaneously drives the system and implicitly defines the dissipative equations of motion required to obtain it. We address this problem using time-dependent constants of motion (invariants)\cite{kaushal1993dynamical}, which uniquely determine the Lindblad jump operators in the corresponding GKLS master equation \cite{dann2021quantum}. This construction yields a self-consistent, thermodynamical framework that faithfully captures both the explicit and implicit time dependence of the controlled open-system dynamics.

With these tools at hand, we used optimal control theory to design gates that are resilient to thermal noise and compared their behavior with that of earlier controller-generated phase- and amplitude-noise sources \cite{aroch2024mitigating}.

For a single qubit with one ancilla (indirect control), a family of pulse solutions was obtained that generates the desired unitary; under thermal noise, this topology is feasible yet intrinsically more complex, with narrower convergence windows and higher residual infidelity than the direct control reference. Allowing additional direct control of the logical transition substantially mitigates the remaining thermal channels. With two or three ancillas, the effective sensitivity to thermal noise is reduced; however, locating isolated, high-fidelity operating points in the noiseless limit remains challenging. 

As a direct-control baseline, a two-qubit controlled-$iX$ (C-iX) gate exhibits the expected temperature-dependent degradation, with slopes determined by the balance between control amplitude and thermal exposure time (see Sec.~\ref{sec:2q-CiX}).

An analysis in Liouville space clarifies the underlying control mechanism. By propagating a full operator basis and reconstructing the CPTP map $\boldsymbol{\Lambda}(\tau)$, we find that optimal control effectively carves out a decoherence-resilient working subspace associated with the logical gate \cite{lidar1998decoherence}. On this reduced subspace, the subsystem purity of the map increases under optimization, indicating that the implemented C-iX transformation becomes more unitary-like, even though the global evolution remains dissipative and the system experiences a net energy loss to the bath. The control field thus rearranges the effective Liouvillian space seen by the computational degrees of freedom, trading global energy relaxation for enhanced subsystem purity of the map and gate fidelity on the relevant subspace.

These findings highlight that mitigation strategies depend on the dominant noise: variance-minimizing, low-amplitude solutions are beneficial for controller noise, whereas thermal relaxation favors shorter exposure or protected pathways in the enlarged Hilbert space. The results provide concrete placements for expanding datasets (single-qubit + 1–3 ancillas; two-qubit C-iX vs.\ temperature) and for benchmarking future control designs under realistic dissipation.

\section*{Acknowledgment}
We gratefully acknowledge the support of the Israel Science Foundation
(grant nos.~510/17 and 526/21).
We also thank Prof.~Roi Baer and Prof.~Raam Uzdin for insightful and
stimulating discussions.

\appendix
\section{Appendix: Numerical methods}
\label{sec:appendix}

\subsection{Vectorizing Liouville Space}
\label{sec:veccing}

Propagators or solvers of the dynamical equations of motion approximate the system’s evolution by expanding the solution in a polynomial series \cite{kosloff1994propagation}. The fundamental computational step in these methods is the matrix-vector multiplication. To apply such propagators to the present open-system dynamics formulated in Liouville space, the numerical scheme must be adapted accordingly.

The following appendix describes the implementation of superoperators acting on operators. This requires vectorizing Liouville space, thereby enabling the use of standard matrix–vector operations. The vectorization procedure is presented both analytically and numerically.

A Hilbert space of operators can be constructed by defining a scalar product on the space of operators. This is equivalent to a linear space of matrices, converting the matrices effectively into vectors $(\rho \rightarrow|\rho\rangle\rangle)$. This is the Fock-Liouville space (FLS) \cite{manzano2020short}. The usual definition of the scalar product of matrices $\phi$ and $\rho$ is defined as $\langle\langle\phi \mid \rho\rangle\rangle \equiv \operatorname{Tr}\left[\phi^{\dagger} \rho\right]$. The Liouville superoperator Eq. (\ref{eq:vonNeumann_isolated}) is now an operator acting on the Hilbert space composed of operators. The main utility of the FLS is to allow the matrix representation of the evolution operator. For example, the qubit density matrix can be expressed in the FLS as
\begin{equation}
   |\rho\rangle\rangle=\left(\begin{array}{c}
\rho_{00} \\
\rho_{01} \\
\rho_{10} \\
\rho_{11}
\end{array}\right). 
\end{equation}
The Liouville von Neumann equation describes the time evolution of a mixed state Eq. (\ref{eq:vonNeumann_isolated}). In vector notation, the Liouvillian superoperators are expressed as a matrix:
\begin{equation}
   {\boldsymbol{\mathcal{L}}}=\left(\begin{array}{cccc}
0 & i \Omega & -i \Omega & 0 \\
i \Omega & i E & 0 & -i \Omega \\
-i \Omega & 0 & -i E & i \Omega \\
0 & -i \Omega & i \Omega & 0
\end{array}\right), 
\end{equation}
Each row is calculated by observing the output of the operation $-i[\hat H, \hat \rho]$ in the computational basis of the density matrix space. The system's time evolution is given by the matrix equation $\left.\frac{d|\rho\rangle\rangle}{d t}=\boldsymbol{\mathcal{L}}|\rho\rangle\right\rangle$, which in matrix notation would be
\begin{equation}
    \left(\begin{array}{l}
\dot{\rho}_{00} \\
\dot{\rho}_{01} \\
\dot{\rho}_{10} \\
\dot{\rho}_{11}
\end{array}\right)=\left(\begin{array}{cccc}
0 & i \Omega & -i \Omega & 0 \\
i \Omega & i E & 0 & -i \Omega \\
-i \Omega & 0 & -i E & i \Omega \\
0 & -i \Omega & i \Omega & 0
\end{array}\right)\left(\begin{array}{l}
\rho_{00} \\
\rho_{01} \\
\rho_{10} \\
\rho_{11}
\end{array}\right) .
\end{equation}
A similar approach is used for the dissipative part ${\boldsymbol{\mathcal{L}}_D}$.

\subsection{Propagation in the Liouville space}
\label{app:prop}

To solve the Liouville–von Neumann equation, achieving high-fidelity control of quantum gates requires highly accurate and efficient numerical propagators. For this purpose, we adapted the semi-global propagator \cite{schaefer2017semi} to operate within the Liouville vector space.
This propagator is explicitly designed to treat generators that are explicitly time dependent and possess complex eigenvalue spectra.

For a driven open system, the propagator is generated by
the Liouvillian ${\boldsymbol{\mathcal{L}}}_t$. In turn, ${\boldsymbol{\mathcal{L}}}_t$ is partitioned into a time-independent and time-dependent generator:
\begin{equation}
\begin{aligned}
\frac{d}{d t} \boldsymbol{\Lambda}(t)
={\boldsymbol{\mathcal{L}}}_t \boldsymbol{\Lambda}(t) =\left(
{\boldsymbol{\mathcal{L}}}_H(t)+{\boldsymbol{\mathcal{L}}}_D(t)\right)\boldsymbol{\Lambda}(t)\\ 
{\boldsymbol{\mathcal{L}}}_H={\boldsymbol{\mathcal{L}}}_{H_0}+{\boldsymbol{\mathcal{L}}}_{H_t}
\end{aligned}
\label{eq:tlio}
\end{equation}
$\boldsymbol{\mathcal{L}}_H(t)$ is the generator of the unitary part of the dynamics in Liouville space:
\begin{equation}
    \frac{d}{dt}\boldsymbol{\Lambda}_t
    = \boldsymbol{\mathcal{L}}_H(t)\boldsymbol{\Lambda}_t,
    \qquad
    \boldsymbol{\mathcal{L}}_H(t)
    = -\frac{i}{\hbar}
      \big(
        \hat{H}_S(t)\otimes I
        - I\otimes\hat{H}_S^{T}(t)
      \big).
    \label{eq:liouvillian_unitary_isolated}
\end{equation}
and can be decomposed into time-independent and time-dependent components. The dissipative generator $\boldsymbol{\mathcal{L}}_D(t)$ implicitly describes the effect of the environment and is also time-dependent to comply with the varying Hamiltonian Eq. (\ref{eq:NAME-GKLS}). 

For a time-independent Lindbladian ${\boldsymbol{\mathcal{L}}}_0$ the formal solution of the dynamics $\frac{d}{dt}\boldsymbol{\Lambda}(t)= {\boldsymbol{\mathcal{L}}}_0 \boldsymbol{\Lambda}(t) $,  the propagator  becomes:
\begin{equation}
    \boldsymbol{\Lambda}(t)=e^{{\boldsymbol{\mathcal{L}}}_0 t}
    \label{time-independent Lindbladian-2}
\end{equation}
with the initial conditions $\boldsymbol{\Lambda}(0)={\cal I}$. We then assume that the Liouvillian can be partitioned into a time-dependent and time-independent part ${\boldsymbol{\mathcal{L}}}={\boldsymbol{\mathcal{L}}}_0+{\boldsymbol{\mathcal{L}}}_t$, a formal solution of Eq. (\ref{eq:tlio}) can be formulated as an integral equation:
\begin{equation}
\boldsymbol{\Lambda}(t)=e^{{\boldsymbol{\boldsymbol{\mathcal{L}}}}_0 t} + \int_0^t e^{{\boldsymbol{\mathcal{L}}}_0 (t-\tau)}{\boldsymbol{\mathcal{L}}}_t \boldsymbol{\Lambda}(\tau) d \tau
\label{integral-solution}
\end{equation}
Eq. (\ref{integral-solution}) will form the basis for the numerical approximation. 

In typical control problems, $\boldsymbol{\mathcal{L}}$ varies considerably with time. Therefore, the total evolution is effectively broken into finite time steps of size $\Delta t$. Then, one can concatenate the propagators and obtain the total evolution from $t=0$ to $t=\tau$ by
\begin{equation}
\boldsymbol{\Lambda}(\tau) \approx \prod_{j=1}^{N_t} {\mathcal Q }_j(\Delta t) 
\label{Prudoct-rule-1}
\end{equation}
where ${\mathcal Q}_j(\Delta t)$ is the propagator for $t$ to $t+\Delta t$ and $t=j \Delta t$. 
A direct approximation assumes that ${\boldsymbol{\mathcal{L}}}_t$ is time-independent within a time step, then
\begin{equation}
    {\mathcal Q_j} \approx e^{ {\boldsymbol{\mathcal{L}}}_j \Delta t}
    \label{eq:gprop}
\end{equation}
where ${\boldsymbol{\mathcal{L}}}_j = {\boldsymbol{\mathcal{L}}}(t+\Delta t/2)$. Sampling ${\boldsymbol{\mathcal{L}}}$ in the middle of the time step leads to second-order accuracy in $\Delta t$.

A numerical method to solve Eq.(\ref{eq:gprop}) is based on expanding the exponent or any analytic function in a polynomial series in the matrix ${\boldsymbol{\mathcal{L}}}_j$:
\begin{equation}
 \boldsymbol{{\mathcal Q_j}}(t+\Delta t) \approx \sum_{n=0}^{K-1} a_n(t+\Delta t) P_n\left(\boldsymbol{\mathcal{L}}_j\right) \boldsymbol{{\mathcal Q_j}}(t) 
 \label{Polynomial_expention}
\end{equation}
where $P_n(\boldsymbol{\mathcal{L}}_j)$ is a polynomial of degree $n$, and $a_n(t+\Delta t)$ is the corresponding expansion coefficient in the interval $t,t+\Delta t$. This requires choosing the set of expansion polynomials $P_n(\boldsymbol{\mathcal{L}}_j)$ and the corresponding coefficients $a_n$  \cite{berman1991time}. 
The expansion (\ref{Polynomial_expention}) has to be accurate in the eigenvalue domain of $\boldsymbol{\mathcal{L}}_j$ so that the form (\ref{Polynomial_expention}) will converge for the representation of $ {\mathcal G_j} $. Successive matrix-vector multiplications generate this series of polynomials at Eq. (\ref{Polynomial_expention}). It scales as {$O\left(KN^2\right)$}, and the computational effort can be reduced even further. For sparse superoperators, the matrix-vector operation can be replaced by an equivalent semi-linear scaling with {$\sim KO(N)$} \cite{kosloff1983fourier}. 

An immediate question concerns the choice of the expansion polynomials $P_n$. In general, one seeks a polynomial basis that achieves the fastest convergence.
An orthogonal set of polynomials is the first step for fast convergence \cite{kosloff1994propagation}.

An efficient implementation can be done recursively. The Chebyshev and Newton interpolation polynomials are two orthogonal expansion series using $P_n(\boldsymbol{\mathcal{L}}_j)$. When the Liouvillian is non-Hermitian, the eigenvalue domain becomes complex, and the Chebyshev approach is no longer appropriate. In this case, the Newton or Arnoldi approach should be adopted \cite{kosloff1994propagation,arnoldi1951principle,lehoucq1996deflation}. 

Note that in Eq. (\ref{time-independent Lindbladian}), only the coefficients $a_n(t)$ are time-dependent. The solution at intermediate time points can be obtained by computing the coefficients at those points with negligible additional computational effort. 

Quantum gate control requires exceptionally high accuracy. The convergence of Eq. (\ref{Prudoct-rule}) with a piecewise constant $\boldsymbol{\mathcal{L}}_j$ is slow, leading to an extensive numerical effort. To achieve faster convergence, we must consider time ordering within the time step $\Delta t$. To overcome the problem of time ordering, we will combine the polynomial solution of Eq. (\ref{time-independent Lindbladian}) and the integral equation formal solution (\ref{integral-solution}). In Eq. (\ref{integral-solution}), the free propagator appears both as a complementary term and in the integrand. The solution of the integral equation requires an iterative approach since $\boldsymbol{\Lambda}(\tau)$ also appears in the integrand. This is done by extrapolating the solution from one time step to the next, from $t$ to $t+dt$. The integral in the formal solution Eq. (\ref{integral-solution}) is reformulated employing an inhomogeneous source term:
\begin{equation}
    \frac{d {\mathcal Q }_j(t)}{d t}=\boldsymbol{\mathcal{L}}_j {\mathcal Q }_j(t)+\overrightarrow{\mathbf{s}}(t)
     \label{source-term}
\end{equation}
The source term will represent the time-dependent/nonlinear part of the dynamics. Treating Eq. (\ref{source-term}) as an inhomogeneous ODE will give rise to a formal solution to the time-dependent problem.

We can write the solution to Eq. (\ref{time-independent Lindbladian}):
\begin{equation}
    \begin{aligned}
\tilde{\boldsymbol{\mathcal Q }}_j(t+\Delta t) & =\tilde{ \boldsymbol{ \mathcal Q}}_j(t)  +\int_t^{t+\Delta t }\tilde{ \boldsymbol{ \mathcal Q}}_j(t-\tau) \overrightarrow{\mathbf{s}}(\tau) d \tau \\
 =&\exp \left(\boldsymbol{\boldsymbol{\mathcal{L}}}_j t\right) +
 \int_t^{t+\Delta t} \exp \left[\boldsymbol{\boldsymbol{\boldsymbol{\mathcal{L}}}}_j(t-\tau)\right] \overrightarrow{\mathbf{s}}(\tau) d \tau \\
 =&\exp \left(\boldsymbol{\mathcal{L}}_j t\right)  
+ \exp \left(\mathcal{L}_j t\right) \int_t^{t+\Delta t} \exp \left(-\boldsymbol{\mathcal{L}}_j \tau\right) \overrightarrow{\mathbf{s}}(\tau) d \tau&
\end{aligned}
\label{Dunhamel-principle}
\end{equation}
Where $\tilde{ \boldsymbol{ \mathcal G}}_j$ is defined as the time-independent propagator by the vec-ing procedure.
The source term $\overrightarrow{\mathbf{s}}(\tau)$ is expanded as a time-dependent polynomial to solve for the integral analytically.
\begin{equation}
    \overrightarrow{\mathbf{s}}(\tau)=\sum_{n=0}^{M-1} \frac{\tau^n}{n !} \overrightarrow{\mathbf{s}}_n
    \label{source-term-polynomial}
\end{equation}
This expansion allows us to solve the integral in Eq. (\ref{Dunhamel-principle}) formally.
\begin{align*}
  \int dt e^{a t} t^m/m ! = \sum_{n=1}^m  e^{a t} t^{n-m}/a^n(n-m)!.  
\end{align*}
The problem is now shifted to obtaining the expansion coefficients $\overrightarrow{\mathbf{s}}_n$. The task is obtained by approximating $\overrightarrow{\boldsymbol{s}}(t)$ by an orthogonal polynomial in the time interval. We choose a Chebyshev expansion.
\begin{equation}
    \overrightarrow{\mathbf{s}}(t) \approx \sum_{n=0}^{M-1} \overrightarrow{\mathbf{b}}_n {\it  T}_n(t)
\label{eq:monomial}
\end{equation}
where the coefficients $\overrightarrow{\mathbf{b}}_n$ are computed by a scalar product of the ${\it T}_n(t)$ with  $\overrightarrow{\mathbf{s}}(t)$. Approximating the coefficients using Chebyshev sampling points in the time interval $\Delta t$.

The coefficients $\overrightarrow{\mathbf{s}}_n$ are calculated by relating the polynomial Eq. (\ref{source-term-polynomial}) to the Chebyshev expansion. This source term is inserted into the integral  Eq. (\ref{Dunhamel-principle}), leading to a numerical approximation to the solution of the TDLE. The addition of the source term into the dynamics gives rise to an analytical solution for the last term in Eq. (\ref{Dunhamel-principle}), presented here on the RHS of Eq.  (\ref{Recortion-integral})
\begin{equation}
    J_{m+1}(\mathcal{L}_j, t) \equiv \int_t^{t+\Delta t} \exp (-\boldsymbol{\mathcal{L}}_j \tau) \tau^m d \tau, \quad m=0,1, \ldots
    \label{Recortion-integral}
\end{equation}
With the recursion relations:
\begin{equation}
\begin{aligned}
      J_m(\boldsymbol{\mathcal{L}}_j, t)=-\frac{\exp (-\boldsymbol{\boldsymbol{\mathcal{L}}}_j t) t^{m-1}}{\boldsymbol{\mathcal{L}}_j}+\frac{m-1}{\boldsymbol{\mathcal{L}}_j} J_{m-1}(\boldsymbol{\mathcal{L}}_j, t), \\ \quad m=2,3, \ldots  
\end{aligned}
\end{equation}
where
\begin{align}
    J_1(\boldsymbol{\mathcal{L}}_j, t) \equiv \int_t^{t+\Delta t} \exp (-\boldsymbol{\mathcal{L}}_j \tau) d \tau=\frac{1-\exp (-\boldsymbol{\mathcal{L}}_j t)}{\boldsymbol{\mathcal{L}}_j}.
\end{align}
Plugging Eq. (\ref{source-term-polynomial}) into this formulation leads to the following:
\begin{equation}
\begin{aligned}
     \exp (\boldsymbol{\mathcal{L}}_j,t) \sum_{n=0}^{M-1} \frac{1}{n !} \int_0^t \exp (-\boldsymbol{\mathcal{L}}_j \tau) t^n d \tau s_n= \\ \exp (\boldsymbol{\mathcal{L}}_j t) \sum_{n=0}^{M-1} \frac{1}{n !} J_{n+1}(\boldsymbol{\mathcal{L}}_j, t) s_n=\sum_{n=0}^{M-1} f_{n+1}(\boldsymbol{\mathcal{L}}_j, t) s_n   
\end{aligned}
\end{equation}
In Eq. (\ref{eq:tlio}), the Liouvillian is split into explicit time-dependent and approximated time-independent parts. The  same analysis leads to:
\begin{equation}
\begin{aligned}
 \boldsymbol{\mathcal G}(t,t+\Delta t)=\exp (\boldsymbol{\mathcal{L}}_jt)+ \\ \exp ({\boldsymbol{\mathcal{L}}_j} t) \int_t^{t+\Delta t} \exp (-{\boldsymbol{\mathcal{L}}_j} \tau) \overrightarrow{\mathbf{s}}(\boldsymbol{\mathcal G}(\tau), \tau) d \tau   
\end{aligned}
\label{eq:integ}
\end{equation}
Now, we can use these formulations to approximate Eq. (\ref{eq:integ}):
\begin{equation}
\begin{aligned}
    \boldsymbol{\mathcal Q}(t,t+ \Delta t) \approx P_M({\boldsymbol{\mathcal{L}}_j}, (t,t+ \Delta t)) \overrightarrow{S}(t,t+ \Delta t)_M+ \\ \sum_{n=0}^{M-1} \frac{t^n}{n !} \overrightarrow S(t,t+ \Delta t)_n  
\end{aligned}
 \label{Final-TDLE inhomogeneous linear ODE}
\end{equation}
$P_M({\boldsymbol{\mathcal{L}}_j}, (t,t+ \Delta t) \overrightarrow{S} (t,t+ \Delta t)_M$, is approximated by the Arnoldi method (the eigenvalue spectrum of $\boldsymbol{\mathcal{L}}$ is distributed on the complex plane), where 
\begin{align*}
            \overrightarrow{{S(t)}}_M \equiv  \overrightarrow{\mathbf{s}}(t)+\boldsymbol{\mathcal{L}}_t \boldsymbol{\mathcal Q}(t)
\end{align*}
$S(t) $in is computed by expanding it by time in the same form as (\ref{source-term-polynomial}). We have used here the fact that $P_n(\boldsymbol{\mathcal{L}}_{H},t)=\boldsymbol{\mathcal{L}}_{H}^{k-n} P_k(\boldsymbol{\mathcal{L}}_{H}, t)+\sum_{j=n}^{k-1} \frac{t^j}{j !} \boldsymbol{\mathcal{L}}_{H}^{j-n}$ Eq. (\ref{Final-TDLE inhomogeneous linear ODE}) and $\overrightarrow{\mathbf{s}}_j$ include dependence on $\boldsymbol{\mathcal Q}(t)$. It would seem we are back to the same problem. However, it can be done through repetition and refinement. 

First, we guess a solution $\boldsymbol{\Lambda}_g(t)$, within a time step $\Delta t$, and use it in Eq. (\ref{Final-TDLE inhomogeneous linear ODE}) to obtain a new approximate solution. This procedure can be continued until the solution converges with the desired accuracy. The initial guess is extrapolated from the previous step to accelerate convergence.

Three numerical parameters determine the precision of the propagation and the convergence rate:
\begin{itemize}
\item{The size of the time step $\Delta t$.}
\item{Number of Chebyshev sampling points in each time step $M$.}
\item{The size of the Krylov space $K$ corresponds to the basis of the Arnoldi algorithm. It is important to note that this parameter is limited by $Dim\{{\boldsymbol{\mathcal{L}}\}-1}$}
\end{itemize}

Each of those is adjustable by the user to best fit their needs. For example, for the  Hadamard propagator system, we use the following parameters: $\Delta t = 0.1$, $M = 7$, and $K=3$. With these parameters, we achieved an accuracy of $10^{-10}$ for the propagator, which is five orders of magnitude higher than the target transformation's fidelity. For the entangling gate, we adjust the parameters($M=K=9$)  to achieve the same resolution of $10^{-10}$.

\subsection{{GRAPE based OCT}}
\label{sec:grape-app}

Another established OCT scheme is GRAPE. It has been shown to lead to
high-fidelity solutions \cite{khaneja2005optimal,goodwin2016modified,petruhanov2022optimal,petruhanov2023grape}.
The starting point is the control Hamiltonian Eq.~(\ref{eq:Hcont}).
Nevertheless, the GRAPE control fields differ significantly from the ones
obtained by the Krotov scheme \cite{palao2003optimal}. GRAPE treats the
control amplitudes as optimization variables and uses gradients of the
objective with respect to the discretized pulse. Let $u_j$ be the control
amplitude on interval $j$. The channel is written as a product of
short-time propagators,
\begin{equation}
\Phi(T)=\boldsymbol{\Lambda}(T) \approx
\prod_{j=1}^{N_t} {\mathcal Q}_j(\Delta t).
\end{equation}

The corresponding derivative is
\begin{equation}
    \frac{\partial {\Lambda}(T)}{\partial u_j}
    =
    {\cal Q}_{N_t}\cdots{\cal Q}_{j+1}
    \frac{\partial{\cal Q}_{j}}{\partial u_j}
    {\cal Q}_{j-1}\cdots{\cal Q}_{1}.
    \label{eq:grape_derivative-2}
\end{equation}
The key numerical issue is the evaluation of
$\partial{\cal Q}_{j}/\partial u_j$. Since the propagation is performed
with an Arnoldi--Chebyshev semi-global scheme, the derivative is evaluated
consistently with the same equation of motion rather than by assuming a
simple first-order exponential derivative. In practice, the matched
sensitivity equation is propagated together with the state,
\begin{align}
    \frac{d}{dt}\left(\frac{\partial{\Phi}}{\partial u_j}\right)
    &=
    {\cal L}(t)\frac{\partial{\Phi}}{\partial u_j}
    +
    \frac{\partial{\cal L}(t)}{\partial u_j}{\Phi}(t).
    \label{eq:sensitivity-1}
\end{align}
This provides a gradient compatible with the chosen propagation scheme and
with the discretization used for the optimized control field.

\begin{figure}[t]
    \hspace{-2cm}
    \includegraphics[width=0.85\linewidth,angle=-90]{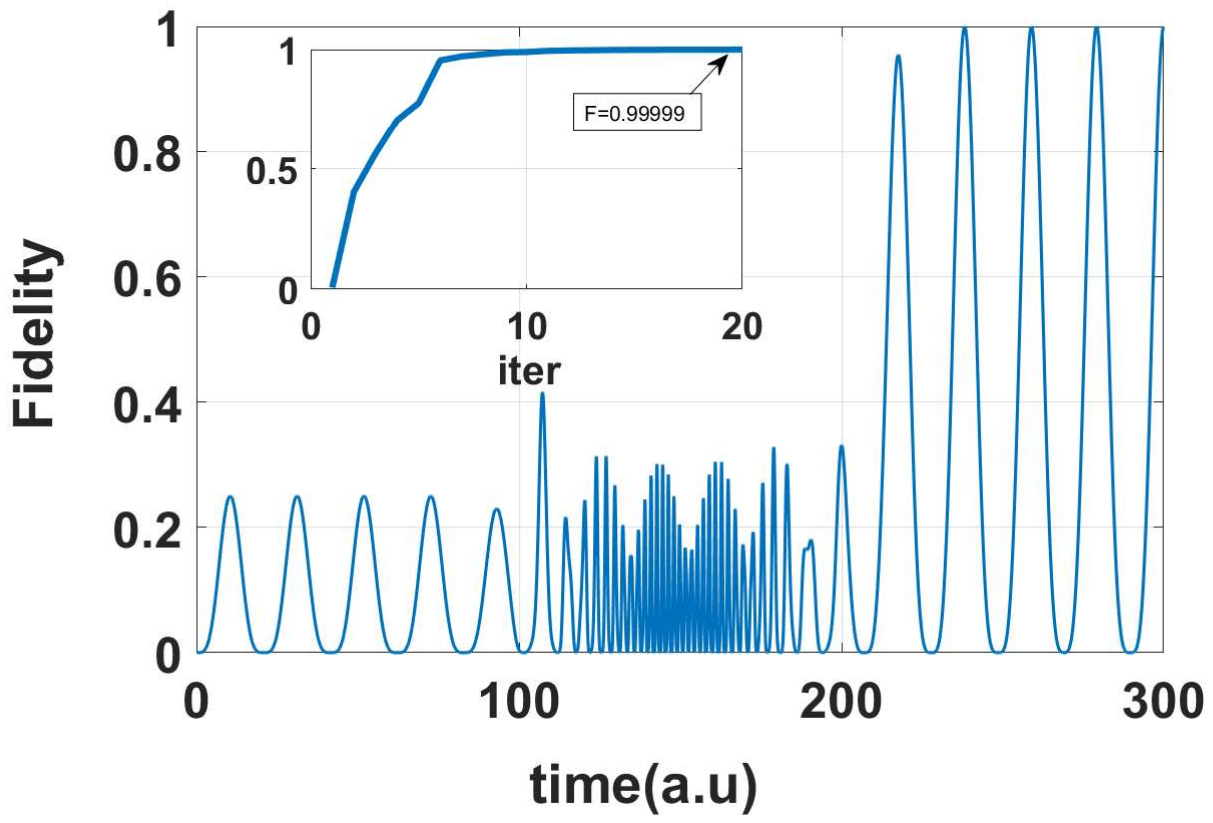}
    \vspace{-2cm}
    \caption{
    Fidelity obtained with the GRAPE optimization. The figure shows the formation of the gate during the propagation (main) and the convergence of the optimization (inset). The monotonic improvement of the optimized objective
    demonstrates that the discretized control amplitudes provide effective
    search directions, while the time-dependent fidelity at the final iteration shows how the target transformation is built over the control window.}
    \label{fig:grape_fidelity}
\end{figure}

The GRAPE calculations were used not only for comparison with the Krotov updates but also as an independent means to enlarge the set of optimized control fields. In this way we obtained many additional high-fidelity solutions and useful initial guesses for the cases considered here, including both single-qubit and two-qubit gate systems. A representative GRAPE result is shown in Fig.~\ref{fig:grape_fidelity}. The availability of several optimized pulses is important because it reduces the possibility
that the observed behavior is tied to a single rare solution or to a particular initial guess.

We also observed indications that the GRAPE solutions can be sensitive to the noise model and to the dissipative parameters. This sensitivity will affect both the convergence path and the final optimized control field. However, a systematic separation between numerical sensitivity, dependence on the initial guess, and genuine physical noise response requires a broader parameter study. We therefore leave a detailed investigation of the
noise-dependent GRAPE landscape for future research.

\section{Tables}

\subsection{Summary of symbols (NAME-GKLS)}
\begin{center}
\captionof{table}{Summary of symbols and definitions used in the NAME-based GKLS formulation.}
\label{tab:variables}

\begin{tabular}{ll}
Symbol & Description \\ \hline
$\hat A_i(t)$ & Time-dependent invariant operators (diagonal sector) \\
$\hat F_{ij}(t)$ & Jump operators connecting instantaneous eigenmodes $j \to i$ \\
$\boldsymbol{\mathcal U}(t)$ & Free evolution super-operator in Liouville space \\
$\phi_{ij}(t)$ & Accumulated dynamical phase,
$\phi_{ij}(t)=\int_0^t \omega_{ij}(t')dt'$ \\
$\omega_{ij}(t)$ & Instantaneous transition (Bohr) frequencies \\
$N$ & Hilbert-space dimension of the system \\
$J(\omega)$ & Bath spectral density evaluated at frequency $\omega$ - Ohmic bath\\
$n_T(\omega)$ & Bose--Einstein occupation,
$n_T(\omega)=(e^{\omega/T}-1)^{-1}$ \\
$\Gamma_{i\leftarrow j}^{\uparrow}(t)$ & Thermal excitation rate from $j$ to $i$ \\
$\Gamma_{j\leftarrow i}^{\downarrow}(t)$ & Thermal relaxation rate from $i$ to $j$ \\
$\gamma$ & Overall system--bath coupling strength \\
$T$ & Bath temperature (units $k_B=\hbar=1$) \\
$\mathcal L_D(t)$ & Time-local GKLS dissipator in the NAME basis \\
$\bullet$ & Placeholder for the density operator argument \\
\end{tabular}

\end{center}

\vspace{1.2em}

\subsection{Simulation parameters for the bare Hamiltonian $H_0$}
\begin{center}
\captionof{table}{Parameters defining the bare Hamiltonian $H_0$, in Hilbert space, used to generate the instantaneous energy levels for each simulated device.
Energies are given in frequency units, and we set $\hbar=1$.}
\label{tab:H0params}
\begin{tabular}{llll}
Device & $N$ & Bare energies / splittings \\ \hline
1Q + 1A &
$N=3$ &
$E_0=0,\;E_1=0.0105$ 
$E_{a1}=0.074$ \\
1Q + 2A &
$N=4$ &
$E_0=0,\;E_1=0.0105$, 
$E_{a1}=0.074$, $E_{a2}=0.094$ \\
1Q + 3A &
$N=5$ &
$E_0=0,\;E_1=0.0105$, 
$E_{a1}=0.074$ $E_{a2}=0.094$,  $E_{a3}=0.114$ \\
\\
\hline
\\
2Q gate (+ ancillas if any) &
$N=4$ &
$\omega_{Q1}=0.15,\;\omega_{Q2}=0.45 $\\
\end{tabular}
\end{center}
\clearpage
\bibliographystyle{plainnat}
\bibliography{ref}

\end{document}